\documentclass[preprint,12pt]{elsarticle}

\UseRawInputEncoding
\usepackage{epsfig}
\usepackage{pgfplots}
\pgfplotsset{compat=1.12}
\usepackage{tikz}
\usepackage{hyperref}
\usepackage{amssymb}
\usepackage{algorithm}
\usepackage{algorithmic} 
\usepackage{multirow}
\usepackage{float}
\usepackage{amsmath}
\usepackage{color}
\usepackage{subfigure}
\usepackage{fancyhdr}
\usepackage{epstopdf}
\usepackage{ulem}
\usepackage{bm}
\usepackage{amsopn}
\usepackage{amssymb,amsmath,color,times}
\usepackage{setspace}

\journal{Computers \& Geotecnics}

\begin{document}

\doublespacing

\begin{frontmatter}

\title{Nonlinear variation of bedload thickness with fluid flow rate in laminar shearing flow}

\author{Duo Wang \corref{cor1}}
\ead{duo.wang1@uq.net.au}

\cortext[cor1]{Corresponding author}

\address{NEPU, China}

\begin{abstract}

The movement of subaqueous sediment in laminar shearing flow is numerically investigated by the coupled lattice Boltzmann and discrete element methods. First, the numerical method is validated by comparing the phase diagram proposed by Ouriemi {\it et al.} ({\it J. Fluid Mech}., vol. 636, 2009, pp. 321-336). Second, a detailed study on sediment movement is performed for sediment with varying solid volume fractions, and a nonlinear relationship between the normalised thickness of the mobile layer and the normalised fluid flow rate is observed for a densely-packed sediment. Third, an independent investigation on the effective viscosity and friction coefficient of the sediment under different fluid flow rates is conducted in a shear cell; and substitution of these two critical parameters into a theoretical expression proposed by Aussillous {\it et al.} ({\it J. Fluid Mech}., vol. 736, 2013, pp. 594-615) provides consistent predictions of bedload thickness with the simulation results of sediment movement. Therefore, we conclude that the non-Newtonian behaviour of densely-packed sediment leads to the nonlinear relationship between the normalised thickness of the mobile layer and the normalised fluid flow rate.

\end{abstract}

\begin{keyword}

sediment movement \sep lattice Boltzmann method \sep discrete element method \sep partially saturated method \sep fluid-particle interactions

\end{keyword}

\end{frontmatter}

\section{Introduction}
\label{sec:introduction}

The movement of a subaqueous sediment exercises a strong influence in a broad range of engineering problems and natural phenomena, such as river morphology in environmental engineering, dune formation in civil and coastal engineering and hydraulic fracturing in mining engineering. For an initially flat sediment which is often modelled as a subaqueous sphere pack, the shearing effect of the upper fluid gives rise to the interfacial instability. When the hydrodynamic force exceeds certain threshold, an erosion-deposition process occurs to the top-layer particles of the sediment first. Continuous fluid--particle and particle--particle interactions set more particles in transport, which eventually leads to the wave-like sediment shapes under certain flow conditions. Although studies have been carried out for over a century, the multi-physics and multi-scale nature of the problem still makes it challenging to fully characterise the sediment movement. For instance, it is usually difficult to get access to the velocity profile and solid volume fraction inside the densely-packed sediment in experiments, while a large portion of the modelling approaches cannot resolve individual particle motion \cite{Vowinckel2021}. Therefore, an efficient particle-resolved modelling approach as well as an improved understanding of the transport mechanism is urgently desired.

Two major approaches can be found in the literature within the context of theoretical modelling of sediment transport. The first model is termed as the erosion-deposition model which was originally proposed by Charru \& Mouilleron-Arnould \cite{Charru2002} through a theoretical analysis under the Couette flow regime. In this step-wise approach, the shear stress exerted by the fluid flow on a fixed wavy bottom was first calculated. The particle transport rate under such shear stress was thus calculated from the viscous resuspension theory \cite{Leighton1986} or the erosion-deposition model \cite{Charru2006a,Charru2006b}, and then used for wave velocity and growth rate predictions. Although successfully captured the initial motion of the particles on the sediment surface, this model has certain limitations as it only considers the motion of the top particle monolayer. The second approach is the continuum model which was fulfilled by Ouriemi et al. \cite{Ouriemi2009a} by the use of a two-phase model. Newtonian rheology was assigned to the fluid phase, and frictional rheology consisting of the Einstein effective viscosity and the Coulomb friction was adopted to model the particle phase. Aussillous et al. \cite{Aussillous2013} later found that such combination was able to recover the experimental results of particle flux, but failed to match the bedload thickness and velocity profiles. Conversely, good agreements with the experiments were obtained by using a granular frictional rheology with a shear-rate-dependent friction coefficient \cite{Cassar2005}. It was a crucial step forward made by Aussillous et al. to take the shear-rate-dependent property of the sediment into consideration. Nevertheless, the effective viscosity was assumed to be constant for simplicity, which is to some extent coarse to fully characterise the sediment movement.

In the present study, we aim to investigate the sediment movement mechanism in laminar flow regime via direct numerical simulation (DNS). Due to the complexity of the problem, empirical solutions are often not able to comprehensively capture the fundamental physical phenomena exhibited by the fluid--particle systems \cite{Feng2010}. To fully resolve the immersed particles, DNS offers an efficient approach to evaluate the hydrodynamic and inter-particle interactions and to track the motions of particles, without employing empirical or analytical models. A significant body of the available DNS study on sediment transport is presented by Kidanemariam \& Uhlmann \cite{Kidanemariam2014a,Kidanemariam2014b,Kidanemariam2017} and Mazzuoli et al. \cite{Mazzuoli2019,Mazzuoli2020}, who incorporated the immersed boundary (IB) technique into the computational fluid dynamics-discrete element method (CFD-DEM) framework to solve the particulate flow. Notable success has been made by CFD-DEM-IB modelling as it accurately resolves the essential aspects involved in the problem. However, the use of a Boolean-type phase indicator sets the solid volume fraction computation inside the sediment in an approximate fashion.

To ensure sufficient precision for hydrodynamic and inter-particle interactions and solid volume fraction computation, the lattice Boltzmann method (LBM) \cite{Chen1992} and the DEM \cite{Cundall1979} are coupled via the modified partially saturated method (MPSM) \cite{Wang2018} in this work to model sediment movement. To the best of our knowledge, very limited information can be found in the literature towards the sediment movement modelling using the coupled LBM-DEM technique. Hence, a brief introduction of the LBM-DEM-MPSM framework is firstly brought out in \S~\ref{sec:methodology}, followed by a verification study. In \S~\ref{sec:investigation}, the rheological properties of the sediments are investigated in detail. By resolving the non-Newtonian behaviour of the sediments with varying solid volume fractions, we look forward to providing new fundamental insights into the sediment movement mechanism.

\section{Numerical modelling of sediment movement}
\label{sec:methodology}

The modelling method implemented in this study features a hydrodynamic coupling between the LBM and the DEM via the MPSM. A detailed introduction to the numerical approach along with is validation through several simple flow configurations can be found in Ref. \cite{Wang2018}. In this section, we further validate the applicability and accuracy of the LBM-DEM-MPSM framework for sediment movement modelling.

\subsection{The LBM-DEM-MPSM framework}
\label{sec:LBM-DEM-MPSM}
\noindent As an alternative to the conventional CFD approaches, the LBM discretises the fluid domain using an orthogonal grid at the mesoscopic scale. Each fluid node on the grid possesses a certain number of particle density distribution functions, $f_{i}$, which is allowed to transport to its neighbouring nodes or stay at rest. During each time iteration, the distribution functions are updated via the `collision-streaming' process, of which collision redistributes $f_{i}$ following the lattice Boltzmann equation (LBE) and streaming propagates the redistributed $f_{i}$ to their adjacent nodes. Once this process is accomplished, the macroscopic variables of the fluid field are obtained from the moments of $f_{i}$. With its concept based on the kinetic theory, the LBM has been widely applied to solve fluid flow due to its straightforward implementation and computational efficiency. In this study, a two-relaxation-time (TRT) \cite{Ginzburg2008} collision operator is utilised to ensure computational accuracy and efficiency.

In this study, the particles are densely packed and in frequent contact. Under the circumstances, the DEM is coupled to the LBM to solve the particle phase. A linear elastic contact model is implemented to compute the kinematics and mechanical interactions of a contact pair by their contact stiffness, $k$, and small overlap $\delta$. The overall solution procedure during each DEM timestep consists of the global contact search, local interaction resolution, contact force calculation and element velocity and position update. It is worth mentioning that no lubrication correction is added to the contact model, as the sediment is densely-packed and close contacts are dominant \cite{Ouriemi2009a}.

\begin{figure}
\centering
\subfigure[]{
\resizebox{0.48\textwidth}{!}{%
\includegraphics{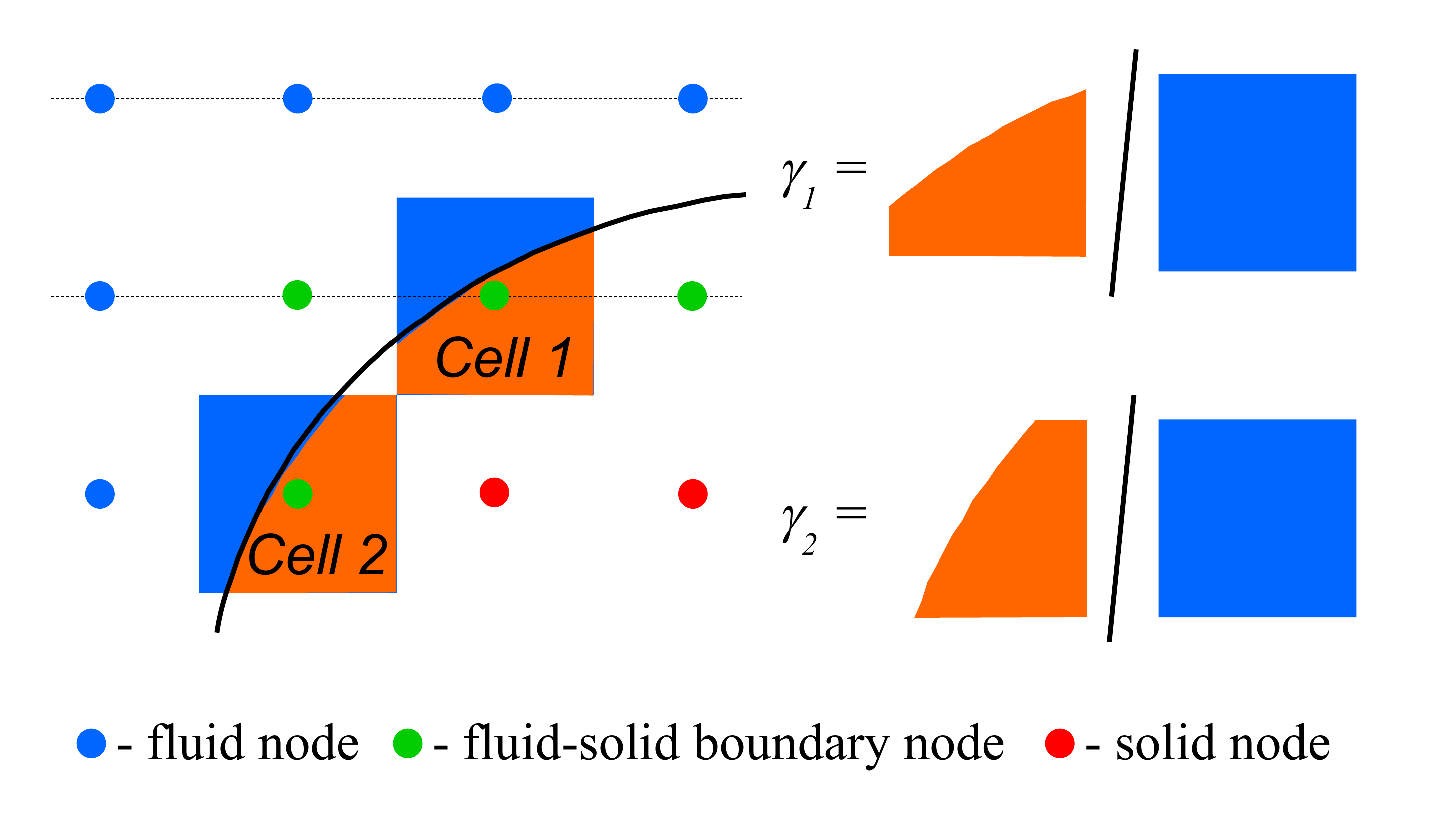}}
\label{fig:PSM}
}
\subfigure[]{
\resizebox{0.35\textwidth}{!}{%
\includegraphics{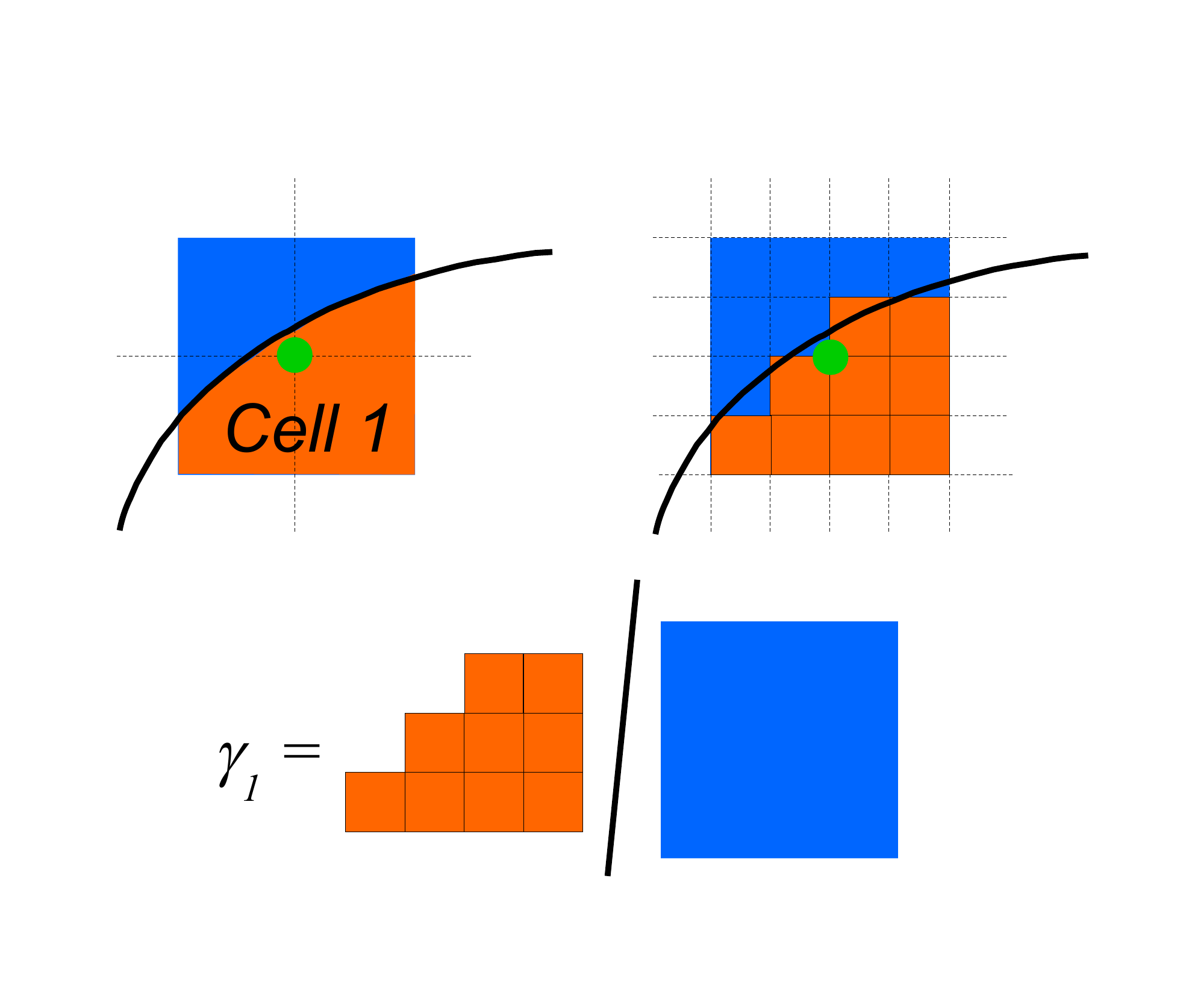}}
\label{fig:Subgrid}
}
\caption{Treatment for fluid--solid boundary cells in the MPSM, showing (a) mapping of the particle boundary on the underlying lattice, and (b) the decomposition of a fluid--solid boundary cell and evaluation of the solid volume fraction, $\gamma$, using the sub-grid technique.}
\label{fig:PSM-combo}
\end{figure}

To fulfil the DNS framework for sediment movement modelling, the MPSM, which is an improvement to the original work by Noble \& Torczynzki \cite{Noble1998}, is employed as the solver for fluid--solid boundaries, see Fig. \ref{fig:PSM}. The introduction of a modified solid weighting function involving the solid volume fraction within a computational cell, $\gamma$, to the LBM improves the description of moving boundaries with which the no-slip boundary is represented. Meanwhile, the use of a sub-grid technique, which is illustrated in Fig. \ref{fig:Subgrid}, improves the mapping precision and minimises the fluctuation in the hydrodynamic force computation. The implementation of the MPSM guarantees a direct, accurate access to the solid volume fraction inside the sediment, which is not straightforward to achieve through experiments or theoretical modellings.

\subsection{Flow configuration and numerical validation}
\label{sec:validation}

\begin{figure}
\centering
\subfigure[]{
\resizebox{0.5\textwidth}{!}{%
\includegraphics{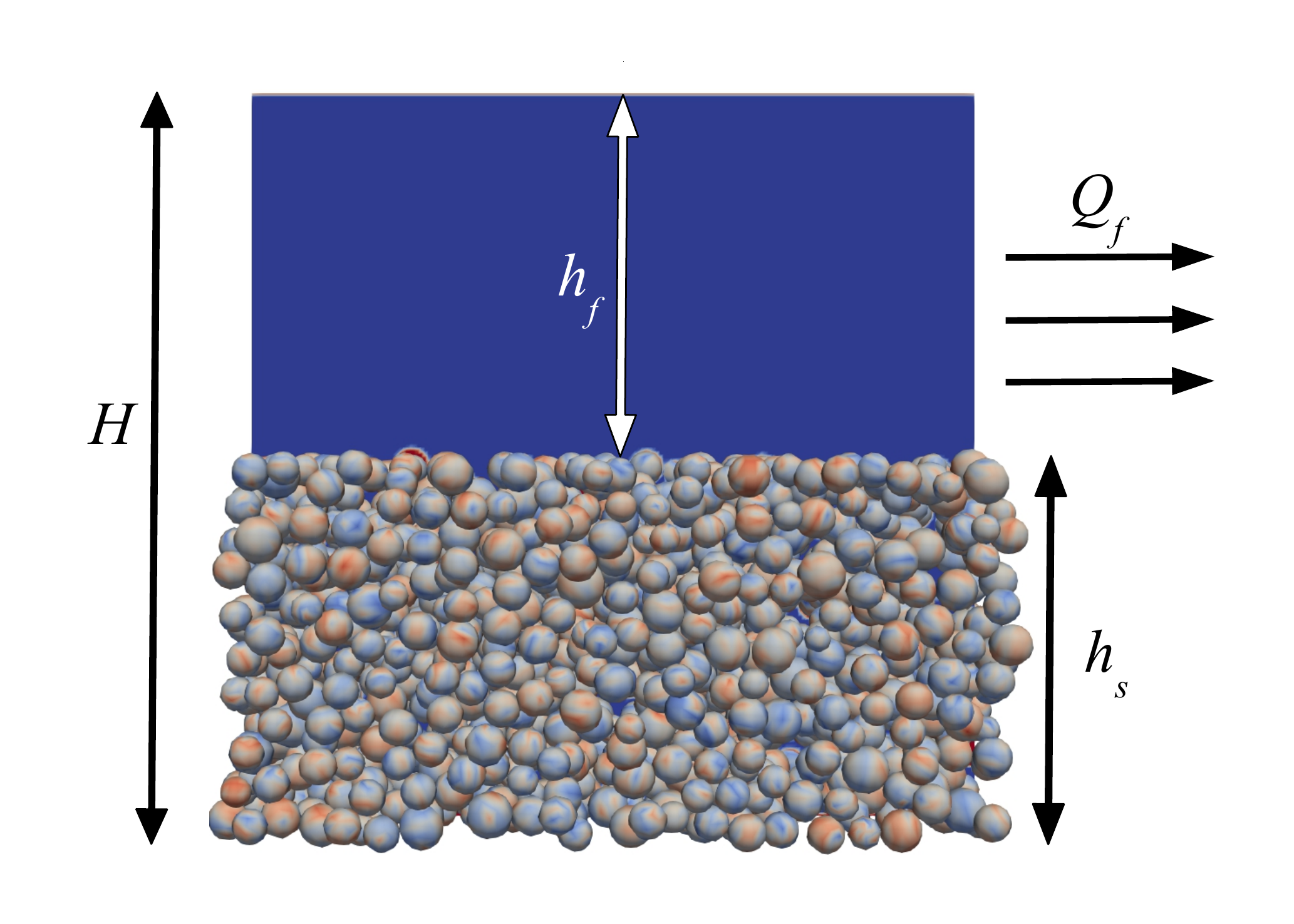}}
\label{fig:sediment-set-up}
}
\subfigure[]{
\resizebox{0.4\textwidth}{!}{%
\includegraphics{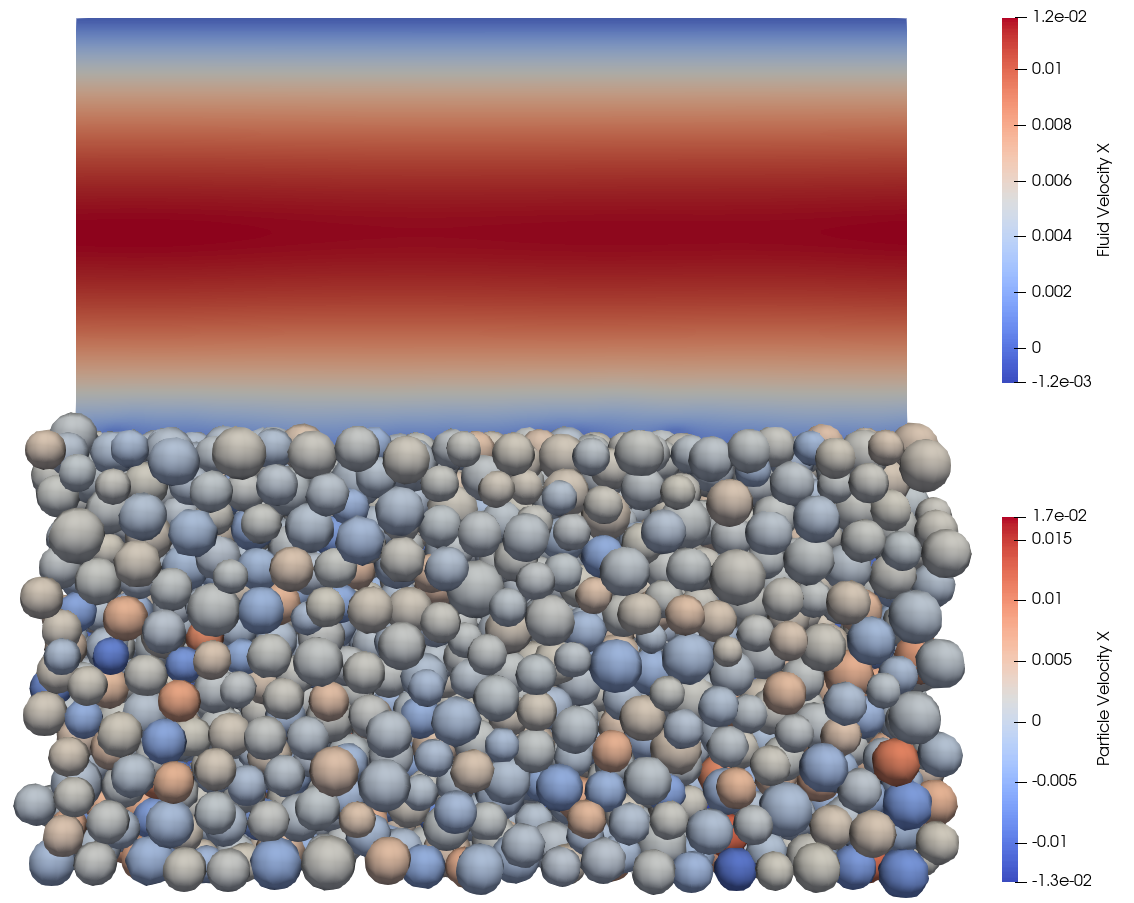}}
\label{fig:sediment-Re0159}
}
\subfigure[]{
\resizebox{0.4\textwidth}{!}{%
\includegraphics{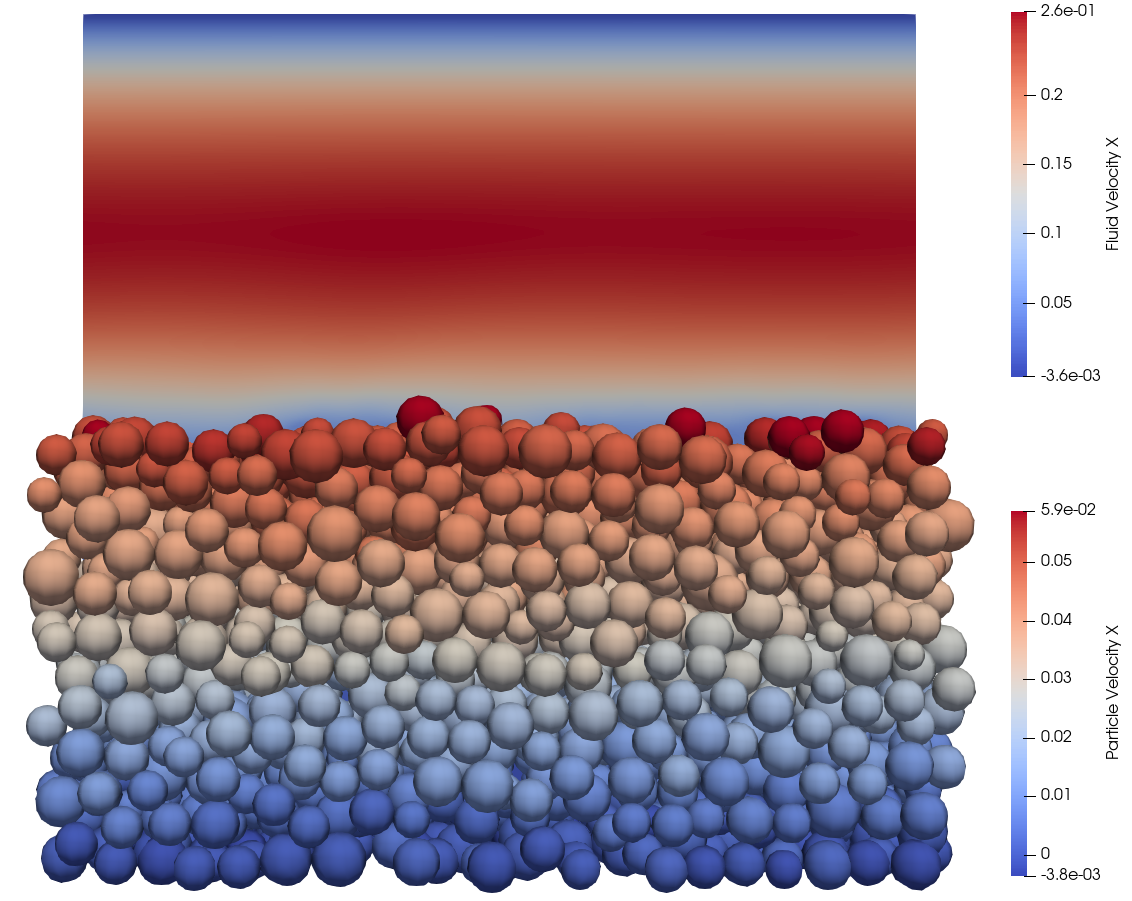}}
\label{fig:sediment-Re3708}
}
\subfigure[]{
\resizebox{0.4\textwidth}{!}{%
\includegraphics{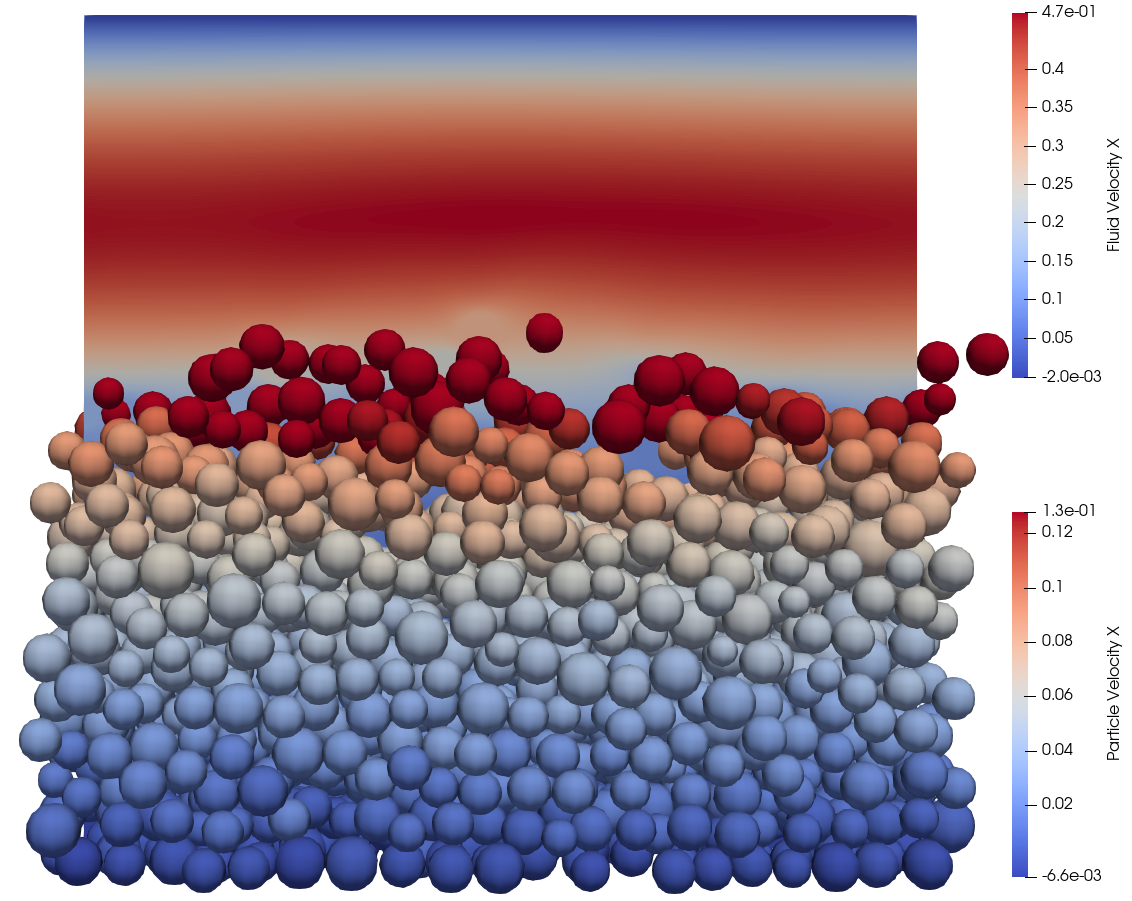}}
\label{fig:sediment-Re6236}
}
\caption{(a) A schematic diagram of the model set-up for sediment transport simulation. (b) Snapshots of the sediment movement modelling at an original $Ga \left( h_{f} / d_{p} \right) ^{2} = 118.36$, showing the `no motion' regime at $Re = 1.59$, (c) the `flat bed motion' regime at $Re = 37.08$ and (d) the `small dunes' regime at $Re = 62.36$. The colour bars represent fluid (upper) and particle (lower) velocities, respectively.}
\label{fig:Sediment-Snapshots}
\end{figure}

We first validate the LBM-DEM-MPSM framework. A schematic diagram of the model construction is shown in Fig. \ref{fig:sediment-set-up}. Randomly packed particles with diameter $d_{p} = 182 \pm 30~\rm{\mu m}$ (equivalent to $100$ to $70$ mesh sand) were distributed in a $3$D channel with dimensions of $0.004~\rm{m} \times 0.002~\rm{m}$ in $L_{x} \times L_{z}$. The particle size followed the normal distribution across the available range. The original depth of the sediment layer $h_{s}$ was $0.002~\rm{m}$, which led to a total particle number of $2236$ to reach a solid volume fraction of $0.45$. The fluid and particle densities were respectively set as $\rho_{f} = 1000~\rm{kg/m^{3}}$ and $\rho_{p} = 2650~\rm{kg/m^{3}}$. The fluid viscosity $\nu$ was $10^{-5}~\rm{m^{2}/s}$. The thickness of the fluid layer $h_{f}$ was varied to achieve different measurements of $Ga \left( h_{f} / d_{p} \right) ^{2}$, in which $Ga$ is the Galileo number. Two definitions of the Galileo number are found in the literature: $Ga = \left( \rho_{p} / \rho_{f} - 1 \right) g d_{p}^{3} / \nu^{2}$ and $Ga = \sqrt{\left( \rho_{p} / \rho_{f} - 1 \right) g d_{p}^{3}} / \nu$ ($g$ is the gravitational acceleration), with one being the square root product of the other. To avoid misunderstandings, the physical properties of the particles were carefully selected in order to make the resultant $Ga \approx 1$.

The LBM relaxation parameter was fixed at $0.8$ for all simulations in this work. The lattice spacing was set as $\Delta x = 20~\rm{\mu m}$ to ensure a sufficient grid resolution for the particles and the friction coefficient of the particles was $0.5$. The no-slip bounce-back condition was applied to the upper boundary and a thin plate with a friction coefficient of $0.5$ was attached to the lower boundary to mimic a rough surface. Periodic boundaries were assigned to the other two directions, thus representing an infinite plane channel. We followed that of Strack \& Cook \cite{Strack2007} to apply a body force, $\textup{G}$, to the upper fluid in the $x$-direction to induce the Poiseuille flow regime over the sediment. The simulations were kept running until both the fluid and particle flow rates reached a steady state. Note that the analysis of this problem features two measurements, one is the aforementioned $Ga \left( h_{f} / d_{p} \right) ^{2}$, the other is the bulk Reynolds number of the channel,

\begin{equation}
Re = Q_{f} / \left( \nu L_{z} \right),
\end{equation}

where $Q_{f}$ is the volumetric fluid flow rate and $L_{z}$ is the channel width.

Figs. \ref{fig:sediment-Re0159} to \ref{fig:sediment-Re6236} present different sediment motion patterns at varying bulk Reynolds number for an initial $Ga \left( h_{f} / d_{p} \right) ^{2} = 118.36$. At $Re = 1.59$, no motion was observed except for some slight particle disturbance due to particle motion in sediment voids. With the increase of fluid flow rate, some of the surface particles exhibited the `rolling, sliding and hopping' behaviour and the `flat bed motion' of the sediment was activated, as indicated by Fig. \ref{fig:sediment-Re3708}. When $Re$ exceeded certain threshold ($Re \approx 37$ predicted by Ouriemi et al. \cite{Ouriemi2009b}, dunes were formed due to the erosion-deposition and continuous interactions of the particles, which accumulated into a sinusoidal bed shape shown in Fig. \ref{fig:sediment-Re6236}.

\begin{figure}
\centering
\resizebox{0.45\textwidth}{!}{
\begin{tikzpicture}

\begin{axis}[%
width=0.45\textwidth,
height=0.4\textwidth,
scale only axis,
xmode=log,
ymode=log,
xmin=10,
xmax=12000,
ymin=0.1,
ymax=100,
xtick={10,1e2,1e3,1e4},
ytick={0.1,1,10,100},
xmajorgrids=true,
xminorgrids=true,
ymajorgrids=true,
grid style=dashed,
xlabel={$Ga \left( h_{f} / d_{p} \right)^{2}$},
label style={font=\large},
ylabel={$Re$},
label style={font=\large},
axis background/.style={fill=white},
legend style={legend pos=south east, legend cell align=left, align=right, draw=black}
]

\addplot [only marks, color=black!30!green, mark=x, mark size=2pt, mark options={solid, black!30!green}]
  table[row sep=crcr]{%
24.01	0.26	\\
24.01	0.42	\\
47.61	0.49	\\
141.61	1.02	\\
141.61	1.63	\\
894.01	9.58	\\
894.01	14.23	\\
};
\addlegendentry{no motion}

\addplot [only marks, color=blue, mark=square*, mark size=2pt, mark options={solid, blue}]
  table[row sep=crcr]{%
24.01	0.84	\\
24.01	3.07	\\
47.61	0.81	\\
47.61	1.41	\\
46.24	2.22	\\
47.61	3.13	\\
46.24	6.58	\\
47.61	12.14	\\
141.61	3.56	\\
141.61	5.69	\\
139.24	12.48	\\
139.24	25.15	\\
888.04	22.38	\\
894.01	33.06	\\
};
\addlegendentry{flat bed motion}

\addplot [only marks, color=red, mark=*, mark size=2pt, mark options={solid, red}]
  table[row sep=crcr]{%
157.5025	62.36	\\
806.56	69.73	\\
1149.21	87.47	\\
};
\addlegendentry{small dunes}

\addplot [dashed, color=black, line width=1.5pt]
  table[row sep=crcr]{%
10	0.14  \\
20000	280  \\
};

\addplot [color=black, line width=1.5pt]
  table[row sep=crcr]{%
10	37  \\
20000	37  \\
};

\addplot [only marks, color=gray, mark=+, mark size=2pt, mark options={solid, gray}]
  table[row sep=crcr]{%
812.84	4.9743	\\
2750.6	17.3792	\\
1155.8128	14.9744	\\
1571.842	19.1932	\\
3173.383	25.8536	\\
2662.4519	26.9	\\
2334.9731	46.4415	\\
};

\addplot [only marks, color=gray, mark=square, mark size=2pt, mark options={solid, gray}]
  table[row sep=crcr]{%
112.4205	0.1	\\
69.4787	0.7926	\\
32.5844	0.884	\\
20.7835	1.0158	\\
27.3529	1.5413	\\
41.0549	1.6855	\\
34.83	2.097	\\
24.2444	1.6855	\\
16.7006	2.6612	\\
11.0078	3.182	\\
51.1594	3.445	\\
59.0065	3.6564	\\
65.1372	4.0784	\\
68.819	4.7334	\\
102.1623	4.9743	\\
160.2318	6.007	\\
53.479	6.0669	\\
26.208	6.4393	\\
31.9327	6.8346	\\
55.2976	11.683	\\
50.671	19.5781	\\
28.96	25.3443	\\
176.9284	9.4843	\\
813.2043	7.1115	\\
804.4961	11.4533	\\
795.9234	15.7366	\\
323.6467	15.1238	\\
353.4184	19.5781	\\
502.0971	19.1932	\\
729.2721	23.8786	\\
1555.0681	25.8526	\\
2334.5164	36.1223	\\
502.3783	39.62	\\
668.2812	42.0521	\\
327.4888	42.8954	\\
257.3105	58.3552	\\
465.3657	56.6427	\\
386.2424	71.8836	\\
};

\addplot [only marks, color=gray, mark=o, mark size=2pt, mark options={solid, gray}]
  table[row sep=crcr]{%
2722.8396	49.2921	\\
1555.387	33.4668	\\
1169.3494	34.8226	\\
970.486	41.6366	\\
918.7959	49.2921	\\
918.7959	58.9375	\\
};

\end{axis}

\end{tikzpicture}

}
\caption{The phase diagram of sediment movement, obtained from our numerical results (filled symbols and crosses) and from Ouriemi et al. \cite{Ouriemi2009b} ( `no motion' $\left( + \right)$, `flat bed motion' $\left( \square \right)$, and `small dunes' $\left( \text{o} \right)$). The dashed line and solid line represent respectively the predicted particle motion threshold and bed instability threshold \cite{Ouriemi2009b}.}
\label{fig:Sediment-Phase}
\end{figure}
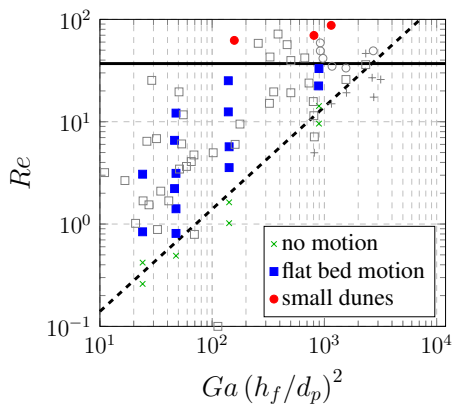

Our numerical results are plotted in the phase diagram in Fig. \ref{fig:Sediment-Phase}, with the experimental results reported by Ouriemi et al. \cite{Ouriemi2009b} incorporated for comparison. It is shown that the simulation results agrees well with the laboratory data in the corresponding sediment motion regime, across the range from the incipient particle motion to dune formation. Different motion regions have been well classified by the data points, which validates our numerical method for sediment movement modelling.

\section{Further investigation of sediment movement}
\label{sec:investigation}

We maintained the same model configuration and parameter settings of \S~\ref{sec:validation} to further study the sediment movement mechanism in laminar flow. The sediment movement was well controlled below the instability threshold, with $Re$ ranging from $0.5$ to $10$ for an original $Ga \left( h_{f} / d_{p} \right) ^{2} = 57.49$. The mobile layer, velocity profile, particle flux and the solid volume faction inside the sediment are presented for sediments with different solid volume fractions. Particularly, the non-Newtonian rheology of the sediments under varying fluid flow rates are investigated in detail via a shear cell model, which provides a means to analyse and predict sediment movement.

\subsection{The mobile layer}
\label{sec:sediment-hm}

Precise predictions for bedload thickness and sediment rheology are of great concern to researchers in this area. In the simulations we vary the applied $\textup{G}$ to investigate the thickness of mobile layer $h_{m}$ under different fluid flow rates. We follow the threshold values of Kidanemariam \& Uhlmann \cite{Kidanemariam2014b} to determine the fluid-sediment interface and $h_{m}$. The upper bound of the mobile layer is at the fluid-sediment interface where the particle solid volume fraction is $\phi = 0.1$, and the lower bound is inside the sediment where the mean particle velocity decreases to 0.5\% of the maximum particle velocity. The scaling of the data follows that of Aussillous et al. \cite{Aussillous2013} by the use of $q^{*}$ which is expressed as,

\begin{equation}
q^{*} = Ga \cdot \nu \cdot \left( h_{f} / d_{p} \right) ^{3}.
\label{eq-qdq*}
\end{equation}

\begin{figure}
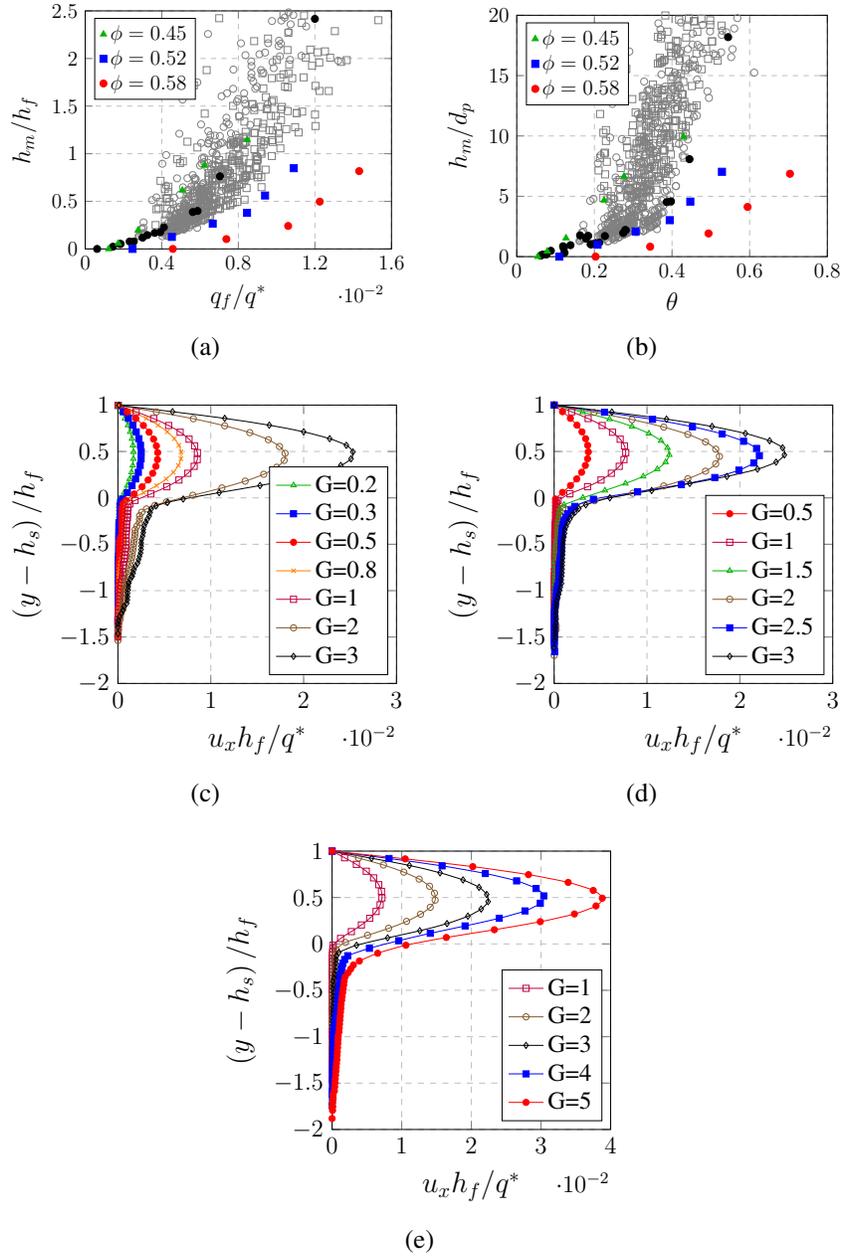

\centering
\subfigure[]{
\resizebox{0.4\textwidth}{!}{%
\input{./Figures/Sediment-hm-q.tex}}
\label{fig:Sediment-hm-q}
}
\subfigure[]{
\resizebox{0.4\textwidth}{!}{%
\input{./Figures/Sediment-hm-theta.tex}}
\label{fig:Sediment-hm-theta}
}\\
\subfigure[]{
\resizebox{0.4\textwidth}{!}{%
\input{./Figures/Interface-svf045.tex}}
\label{fig:Interface-svf045}
}
\subfigure[]{
\resizebox{0.4\textwidth}{!}{%
\begin{tikzpicture}

\begin{axis}[%
width=0.35\textwidth,
height=0.35\textwidth,
scale only axis,
xmin=0,
xmax=0.03,
ymin=-2,
ymax=1,
xtick={0,0.01,0.02,0.03},
ytick={-2,-1.5,-1,-0.5,0,0.5,1},
xmajorgrids=true,
xminorgrids=true,
ymajorgrids=true,
grid style=dashed,
xlabel={$u_{x} h_{f}/ q^{*}$},
label style={font=\large},
ylabel={$\left( y - h_{s} \right) / h_{f}$},
label style={font=\large},
axis background/.style={fill=white},
legend style={legend pos=south east, legend cell align=left, align=right, draw=black}
]

\addplot [smooth, color=red, mark=*, mark size=1.5pt, mark options={solid, red}]
  table[row sep=crcr]{%
1.35E-07	-1.463768116	\\
2.61E-05	-1.389855079	\\
6.23E-05	-1.352898561	\\
8.40E-05	-1.318405808	\\
1.09E-04	-1.281449279	\\
4.16E-05	-1.244492771	\\
6.07E-05	-1.210000008	\\
6.07E-05	-1.1730435	\\
3.87E-05	-1.136086971	\\
8.50E-06	-1.099130442	\\
3.81E-05	-1.064637721	\\
2.66E-05	-1.027681171	\\
5.13E-06	-0.990724663	\\
2.70E-06	-0.9562319	\\
2.70E-06	-0.919275392	\\
2.70E-06	-0.882318884	\\
1.89E-05	-0.847826121	\\
3.17E-05	-0.810869613	\\
3.17E-05	-0.773913063	\\
8.89E-05	-0.736956555	\\
1.10E-04	-0.702463792	\\
8.55E-05	-0.665507326	\\
7.75E-05	-0.628550776	\\
7.75E-05	-0.594058013	\\
7.75E-05	-0.557101463	\\
7.37E-05	-0.520144997	\\
1.28E-04	-0.483188447	\\
1.29E-04	-0.448695684	\\
9.65E-05	-0.411739218	\\
7.28E-05	-0.374782668	\\
7.28E-05	-0.340289905	\\
7.28E-05	-0.303333439	\\
9.36E-05	-0.266376889	\\
1.39E-04	-0.231884126	\\
1.28E-04	-0.194927576	\\
1.28E-04	-0.15797111	\\
1.40E-04	-0.12101456	\\
1.41E-04	-0.086521797	\\
1.41E-04	-0.049565331	\\
1.41E-04	-0.012608781	\\
9.01E-04	0.058840532	\\
1.76E-03	0.132753463	\\
2.43E-03	0.204202777	\\
2.96E-03	0.27565209	\\
3.36E-03	0.34956519	\\
3.60E-03	0.421014335	\\
3.68E-03	0.494927435	\\
3.61E-03	0.566376748	\\
3.39E-03	0.637825893	\\
3.00E-03	0.711738993	\\
2.46E-03	0.783188306	\\
1.75E-03	0.857101237	\\
8.97E-04	0.928550551	\\
4.44E-09	0.999999864	\\
};
\addlegendentry{G=0.5}

\addplot [smooth, color=purple, mark=square, mark size=1.5pt, mark options={solid, purple}]
  table[row sep=crcr]{%
4.63E-07	-1.5	\\
2.99E-06	-1.425000007	\\
1.65E-05	-1.38750001	\\
2.16E-05	-1.352500011	\\
2.16E-05	-1.315000004	\\
2.16E-05	-1.277500018	\\
1.51E-05	-1.242500008	\\
4.16E-05	-1.205000022	\\
3.61E-05	-1.167500015	\\
2.76E-05	-1.130000007	\\
3.95E-05	-1.095000041	\\
3.30E-05	-1.057500012	\\
4.29E-05	-1.020000026	\\
4.29E-05	-0.985000016	\\
4.29E-05	-0.94750003	\\
3.98E-05	-0.910000044	\\
4.72E-05	-0.875000035	\\
2.48E-05	-0.837500049	\\
5.46E-05	-0.80000002	\\
6.33E-05	-0.762500034	\\
6.33E-05	-0.727500024	\\
6.24E-05	-0.690000081	\\
5.92E-05	-0.652500052	\\
6.33E-05	-0.617500043	\\
6.90E-05	-0.580000014	\\
7.39E-05	-0.542500071	\\
7.54E-05	-0.505000042	\\
7.34E-05	-0.470000032	\\
7.58E-05	-0.432500089	\\
7.10E-05	-0.39500006	\\
1.09E-04	-0.360000051	\\
1.42E-04	-0.322500107	\\
1.43E-04	-0.285000079	\\
1.58E-04	-0.250000069	\\
1.97E-04	-0.21250004	\\
2.52E-04	-0.175000097	\\
2.54E-04	-0.137500068	\\
2.55E-04	-0.102500059	\\
2.89E-04	-0.065000116	\\
4.36E-04	-0.027500087	\\
1.70E-03	0.044999952	\\
3.60E-03	0.119999838	\\
5.05E-03	0.192499876	\\
6.16E-03	0.264999915	\\
7.01E-03	0.339999972	\\
7.52E-03	0.41249984	\\
7.71E-03	0.487499897	\\
7.57E-03	0.559999936	\\
7.11E-03	0.632499803	\\
6.29E-03	0.70749986	\\
5.17E-03	0.779999899	\\
3.66E-03	0.854999785	\\
1.88E-03	0.927499823	\\
8.10E-09	0.999999862	\\
};
\addlegendentry{G=1}

\addplot [smooth, color=black!30!green, mark=triangle, mark size=1.5pt, mark options={solid, black!30!green}]
  table[row sep=crcr]{%
6.54E-07	-1.615384615	\\
3.15E-05	-1.536923084	\\
3.38E-05	-1.497692318	\\
3.40E-05	-1.461076935	\\
3.42E-05	-1.421846158	\\
3.45E-05	-1.382615403	\\
3.47E-05	-1.346000008	\\
3.49E-05	-1.306769254	\\
3.52E-05	-1.267538477	\\
3.54E-05	-1.2283077	\\
3.56E-05	-1.19169235	\\
3.59E-05	-1.152461551	\\
3.61E-05	-1.113230796	\\
3.63E-05	-1.076615401	\\
3.66E-05	-1.037384647	\\
3.73E-05	-0.998153892	\\
4.30E-05	-0.961538498	\\
4.90E-05	-0.922307743	\\
5.50E-05	-0.883076944	\\
6.10E-05	-0.843846189	\\
6.66E-05	-0.807230795	\\
7.27E-05	-0.768000085	\\
7.87E-05	-0.728769285	\\
8.43E-05	-0.692153891	\\
9.03E-05	-0.652923091	\\
9.64E-05	-0.613692382	\\
1.02E-04	-0.574461582	\\
1.72E-04	-0.537846188	\\
2.29E-04	-0.498615478	\\
2.63E-04	-0.459384679	\\
3.00E-04	-0.422769284	\\
3.41E-04	-0.383538574	\\
3.45E-04	-0.344307775	\\
3.57E-04	-0.30769238	\\
3.83E-04	-0.268461581	\\
4.21E-04	-0.229230871	\\
5.11E-04	-0.190000072	\\
5.52E-04	-0.153384677	\\
5.96E-04	-0.114153967	\\
1.06E-03	-0.074923168	\\
3.10E-03	0.000923026	\\
5.47E-03	0.079384446	\\
7.68E-03	0.15523064	\\
9.59E-03	0.231076834	\\
1.11E-02	0.309538433	\\
1.20E-02	0.385384448	\\
1.24E-02	0.463846046	\\
1.22E-02	0.53969224	\\
1.15E-02	0.615538255	\\
1.02E-02	0.693999854	\\
8.39E-03	0.769846048	\\
5.96E-03	0.848307467	\\
3.07E-03	0.924153662	\\
3.92E-07	0.999999856	\\
};
\addlegendentry{G=1.5}

\addplot [smooth, color=black!30!brown, mark=o, mark size=1.5pt, mark options={solid, black!30!brown}]
  table[row sep=crcr]{%
3.44E-06	-1.698412698	\\
1.87E-05	-1.617460325	\\
1.87E-05	-1.576984138	\\
1.87E-05	-1.539206361	\\
8.29E-06	-1.498730163	\\
5.07E-05	-1.458253987	\\
2.20E-04	-1.420476199	\\
2.20E-04	-1.380000024	\\
1.44E-05	-1.339523825	\\
9.70E-05	-1.299047627	\\
4.60E-05	-1.261269885	\\
2.29E-05	-1.220793663	\\
2.29E-05	-1.180317488	\\
2.29E-05	-1.1425397	\\
8.57E-05	-1.102063525	\\
1.89E-04	-1.061587349	\\
1.91E-04	-1.023809561	\\
2.32E-04	-0.983333386	\\
2.89E-04	-0.942857164	\\
2.68E-04	-0.902380989	\\
2.39E-04	-0.864603201	\\
2.43E-04	-0.824127072	\\
2.76E-04	-0.78365085	\\
1.70E-04	-0.745873062	\\
6.18E-05	-0.70539684	\\
2.10E-04	-0.664920711	\\
2.63E-04	-0.62444449	\\
1.89E-04	-0.586666701	\\
1.82E-04	-0.546190572	\\
2.14E-04	-0.505714351	\\
2.54E-04	-0.467936563	\\
3.77E-04	-0.427460433	\\
3.79E-04	-0.386984212	\\
3.79E-04	-0.349206424	\\
4.94E-04	-0.308730202	\\
6.13E-04	-0.268254073	\\
6.16E-04	-0.227777852	\\
6.81E-04	-0.190000063	\\
6.81E-04	-0.149523934	\\
9.54E-04	-0.109047713	\\
4.35E-03	-0.030793703	\\
8.47E-03	0.050158555	\\
1.18E-02	0.128412565	\\
1.44E-02	0.206666575	\\
1.63E-02	0.287619018	\\
1.74E-02	0.365872843	\\
1.78E-02	0.446825286	\\
1.74E-02	0.525079296	\\
1.63E-02	0.60333312	\\
1.44E-02	0.684285563	\\
1.18E-02	0.762539573	\\
8.39E-03	0.843491831	\\
4.30E-03	0.921745841	\\
4.29E-08	0.999999851	\\
};
\addlegendentry{G=2}

\addplot [smooth, color=blue, mark=square*, mark size=1.5pt, mark options={solid, blue}]
  table[row sep=crcr]{%
5.67E-05	-1.65625	\\
1.53E-05	-1.576562507	\\
2.63E-06	-1.536718761	\\
2.38E-05	-1.499531262	\\
3.48E-05	-1.459687504	\\
3.52E-05	-1.419843769	\\
3.56E-05	-1.382656259	\\
3.59E-05	-1.342812523	\\
3.63E-05	-1.302968766	\\
3.67E-05	-1.263125008	\\
3.70E-05	-1.225937543	\\
8.38E-05	-1.186093763	\\
1.35E-04	-1.146250027	\\
1.83E-04	-1.109062517	\\
2.34E-04	-1.069218782	\\
2.68E-04	-1.029375047	\\
2.68E-04	-0.992187537	\\
3.55E-04	-0.952343802	\\
4.76E-04	-0.912500021	\\
5.38E-04	-0.872656286	\\
5.41E-04	-0.835468776	\\
5.43E-04	-0.795625086	\\
5.46E-04	-0.755781306	\\
5.48E-04	-0.718593795	\\
6.26E-04	-0.678750015	\\
6.67E-04	-0.638906325	\\
7.45E-04	-0.599062545	\\
7.47E-04	-0.561875034	\\
7.50E-04	-0.522031345	\\
7.64E-04	-0.482187564	\\
7.89E-04	-0.445000054	\\
8.17E-04	-0.405156364	\\
8.44E-04	-0.365312584	\\
8.70E-04	-0.328125073	\\
9.65E-04	-0.288281293	\\
1.07E-03	-0.248437603	\\
1.28E-03	-0.208593823	\\
1.49E-03	-0.171406312	\\
1.83E-03	-0.131562623	\\
2.22E-03	-0.091718842	\\
4.23E-03	-0.014687551	\\
8.99E-03	0.064999828	\\
1.37E-02	0.142031119	\\
1.73E-02	0.21906241	\\
1.99E-02	0.298749971	\\
2.15E-02	0.37578108	\\
2.21E-02	0.455468641	\\
2.18E-02	0.532499932	\\
2.04E-02	0.60953104	\\
1.81E-02	0.689218602	\\
1.49E-02	0.766249892	\\
1.06E-02	0.845937272	\\
5.43E-03	0.922968562	\\
2.45E-08	0.999999853	\\
};
\addlegendentry{G=2.5}

\addplot [smooth, color=black, mark=diamond, mark size=1.5pt, mark options={solid, black}]
  table[row sep=crcr]{%
4.79E-07	-1.615384615	\\
4.00E-05	-1.536923084	\\
1.28E-04	-1.497692318	\\
1.27E-04	-1.461076935	\\
1.27E-04	-1.421846158	\\
9.48E-05	-1.382615403	\\
2.07E-04	-1.346000008	\\
2.11E-04	-1.306769254	\\
1.62E-04	-1.267538477	\\
1.84E-04	-1.2283077	\\
2.08E-04	-1.19169235	\\
2.34E-04	-1.152461551	\\
2.65E-04	-1.113230796	\\
3.87E-04	-1.076615401	\\
5.00E-04	-1.037384647	\\
5.85E-04	-0.998153892	\\
6.70E-04	-0.961538498	\\
7.95E-04	-0.922307743	\\
7.73E-04	-0.883076944	\\
8.25E-04	-0.843846189	\\
8.50E-04	-0.807230795	\\
8.64E-04	-0.768000085	\\
8.77E-04	-0.728769285	\\
8.90E-04	-0.692153891	\\
9.03E-04	-0.652923091	\\
9.17E-04	-0.613692382	\\
9.30E-04	-0.574461582	\\
9.43E-04	-0.537846188	\\
9.56E-04	-0.498615478	\\
9.84E-04	-0.459384679	\\
1.06E-03	-0.422769284	\\
1.12E-03	-0.383538574	\\
1.16E-03	-0.344307775	\\
1.29E-03	-0.30769238	\\
1.45E-03	-0.268461581	\\
1.68E-03	-0.229230871	\\
1.91E-03	-0.190000072	\\
2.27E-03	-0.153384677	\\
2.66E-03	-0.114153967	\\
3.41E-03	-0.074923168	\\
6.00E-03	0.000923026	\\
1.05E-02	0.079384446	\\
1.46E-02	0.15523064	\\
1.86E-02	0.231076834	\\
2.19E-02	0.309538433	\\
2.39E-02	0.385384448	\\
2.48E-02	0.463846046	\\
2.46E-02	0.53969224	\\
2.32E-02	0.615538255	\\
2.06E-02	0.693999854	\\
1.70E-02	0.769846048	\\
1.21E-02	0.848307467	\\
6.22E-03	0.924153662	\\
1.76E-08	0.999999856	\\
};
\addlegendentry{G=3}

\end{axis}

\end{tikzpicture}

}
\label{fig:Interface-svf052}
}
\subfigure[]{
\resizebox{0.4\textwidth}{!}{%
\begin{tikzpicture}

\begin{axis}[%
width=0.35\textwidth,
height=0.35\textwidth,
scale only axis,
xmin=0,
xmax=0.04,
ymin=-2,
ymax=1,
xtick={0,0.01,0.02,0.03,0.04},
ytick={-2,-1.5,-1,-0.5,0,0.5,1},
xmajorgrids=true,
xminorgrids=true,
ymajorgrids=true,
grid style=dashed,
xlabel={$u_{x} h_{f}/ q^{*}$},
label style={font=\large},
ylabel={$\left( y - h_{s} \right) / h_{f}$},
label style={font=\large},
axis background/.style={fill=white},
legend style={legend pos=south east, legend cell align=left, align=right, draw=black}
]

\addplot [smooth, color=purple, mark=square, mark size=1.5pt, mark options={solid, purple}]
  table[row sep=crcr]{%
3.15E-07	-1.463768116	\\
1.89E-05	-1.389855079	\\
2.82E-06	-1.355362326	\\
1.76E-05	-1.318405808	\\
1.76E-05	-1.281449279	\\
1.59E-05	-1.244492771	\\
3.06E-07	-1.210000008	\\
1.53E-06	-1.1730435	\\
5.47E-06	-1.136086971	\\
5.84E-06	-1.099130442	\\
5.90E-06	-1.064637721	\\
1.21E-05	-1.027681171	\\
2.63E-05	-0.990724663	\\
2.63E-05	-0.9562319	\\
2.63E-05	-0.921739179	\\
2.36E-05	-0.882318884	\\
9.35E-06	-0.847826121	\\
2.05E-05	-0.810869613	\\
3.00E-05	-0.773913063	\\
1.94E-05	-0.736956555	\\
5.51E-05	-0.702463792	\\
3.36E-06	-0.665507326	\\
1.67E-05	-0.628550776	\\
2.15E-05	-0.594058013	\\
1.70E-06	-0.557101463	\\
2.02E-05	-0.520144997	\\
6.95E-06	-0.483188447	\\
3.58E-05	-0.448695684	\\
4.94E-05	-0.411739218	\\
3.26E-05	-0.374782668	\\
3.30E-05	-0.340289905	\\
1.62E-05	-0.303333439	\\
5.51E-05	-0.266376889	\\
3.50E-05	-0.231884126	\\
3.96E-05	-0.194927576	\\
2.24E-05	-0.15797111	\\
4.47E-05	-0.12101456	\\
7.54E-05	-0.086521797	\\
1.18E-04	-0.049565331	\\
1.65E-04	-0.012608781	\\
1.22E-03	0.058840532	\\
2.77E-03	0.132753463	\\
4.21E-03	0.204202777	\\
5.40E-03	0.27565209	\\
6.32E-03	0.34956519	\\
6.90E-03	0.421014335	\\
7.16E-03	0.494927435	\\
7.09E-03	0.566376748	\\
6.70E-03	0.637825893	\\
5.96E-03	0.711738993	\\
4.91E-03	0.783188306	\\
3.49E-03	0.857101237	\\
1.80E-03	0.928550551	\\
1.30E-08	0.999999864	\\
};
\addlegendentry{G=1}

\addplot [smooth, color=black!30!brown, mark=o, mark size=1.5pt, mark options={solid, black!30!brown}]
  table[row sep=crcr]{%
1.16E-07	-1.575757576	\\
1.42E-04	-1.498484855	\\
1.40E-04	-1.46242425	\\
1.40E-04	-1.459848495	\\
8.88E-05	-1.42378789	\\
1.16E-04	-1.385151519	\\
2.11E-05	-1.34651517	\\
5.24E-05	-1.310454554	\\
1.16E-04	-1.271818205	\\
1.51E-05	-1.233181833	\\
1.32E-05	-1.194545462	\\
7.31E-05	-1.15848489	\\
6.63E-05	-1.119848497	\\
8.61E-06	-1.081212148	\\
7.58E-06	-1.045151532	\\
1.11E-04	-1.00909096	\\
1.11E-04	-0.967878833	\\
1.11E-04	-0.931818217	\\
2.11E-05	-0.893181868	\\
3.08E-06	-0.854545475	\\
3.08E-06	-0.815909126	\\
3.08E-06	-0.77984851	\\
9.31E-05	-0.741212205	\\
1.29E-04	-0.702575811	\\
4.30E-05	-0.666515195	\\
4.30E-05	-0.627878802	\\
4.30E-05	-0.589242497	\\
2.60E-05	-0.550606104	\\
1.81E-04	-0.514545488	\\
1.42E-04	-0.475909183	\\
8.01E-05	-0.437272789	\\
9.14E-05	-0.401212173	\\
1.37E-04	-0.362575868	\\
1.26E-04	-0.323939475	\\
1.33E-04	-0.287878859	\\
1.47E-04	-0.249242466	\\
1.57E-04	-0.210606161	\\
1.90E-04	-0.171969767	\\
2.81E-04	-0.135909151	\\
4.04E-04	-0.097272846	\\
5.72E-04	-0.058636453	\\
1.92E-03	0.016060556	\\
5.10E-03	0.093333166	\\
8.12E-03	0.168030176	\\
1.07E-02	0.242727185	\\
1.27E-02	0.319999972	\\
1.41E-02	0.394696804	\\
1.48E-02	0.471969591	\\
1.47E-02	0.5466666	\\
1.40E-02	0.621363433	\\
1.25E-02	0.69863622	\\
1.03E-02	0.773333229	\\
7.38E-03	0.850605839	\\
3.80E-03	0.925302848	\\
1.57E-08	0.999999858	\\
};
\addlegendentry{G=2}

\addplot [smooth, color=black, mark=diamond, mark size=1.5pt, mark options={solid, black}]
  table[row sep=crcr]{%
8.83E-07	-1.65625	\\
3.96E-05	-1.576562507	\\
4.02E-05	-1.539375008	\\
4.03E-05	-1.536718761	\\
4.09E-05	-1.499531262	\\
4.15E-05	-1.459687504	\\
4.22E-05	-1.419843769	\\
4.28E-05	-1.382656259	\\
4.35E-05	-1.342812523	\\
4.41E-05	-1.302968766	\\
4.48E-05	-1.263125008	\\
4.54E-05	-1.225937543	\\
4.61E-05	-1.186093763	\\
4.67E-05	-1.146250027	\\
4.73E-05	-1.109062517	\\
4.80E-05	-1.071875052	\\
4.87E-05	-1.029375047	\\
4.93E-05	-0.992187537	\\
4.99E-05	-0.952343802	\\
5.06E-05	-0.912500021	\\
5.12E-05	-0.872656286	\\
5.19E-05	-0.835468776	\\
5.25E-05	-0.795625086	\\
9.66E-05	-0.755781306	\\
1.02E-04	-0.718593795	\\
1.07E-04	-0.678750015	\\
1.74E-04	-0.638906325	\\
1.75E-04	-0.599062545	\\
1.76E-04	-0.561875034	\\
1.76E-04	-0.522031345	\\
1.26E-04	-0.482187564	\\
1.83E-04	-0.445000054	\\
1.31E-04	-0.405156364	\\
2.87E-04	-0.365312584	\\
3.86E-04	-0.328125073	\\
5.29E-04	-0.288281293	\\
5.79E-04	-0.248437603	\\
5.60E-04	-0.208593823	\\
6.03E-04	-0.171406312	\\
6.42E-04	-0.131562623	\\
9.78E-04	-0.091718842	\\
3.51E-03	-0.014687551	\\
7.98E-03	0.064999828	\\
1.26E-02	0.142031119	\\
1.65E-02	0.21906241	\\
1.95E-02	0.298749971	\\
2.14E-02	0.37578108	\\
2.24E-02	0.455468641	\\
2.22E-02	0.532499932	\\
2.11E-02	0.60953104	\\
1.88E-02	0.689218602	\\
1.55E-02	0.766249892	\\
1.11E-02	0.845937272	\\
5.67E-03	0.922968562	\\
1.70E-06	0.999999853	\\
};
\addlegendentry{G=3}

\addplot [smooth, color=blue, mark=square*, mark size=1.5pt, mark options={solid, blue}]
  table[row sep=crcr]{%
2.40E-09	-1.741935484	\\
5.64E-05	-1.659677427	\\
6.22E-05	-1.621290331	\\
6.26E-05	-1.618548398	\\
6.84E-05	-1.580161302	\\
7.46E-05	-1.539032262	\\
8.08E-05	-1.497903245	\\
8.66E-05	-1.459516138	\\
9.28E-05	-1.418387121	\\
9.90E-05	-1.377258081	\\
1.05E-04	-1.33612904	\\
1.11E-04	-1.29774198	\\
1.17E-04	-1.256612916	\\
1.23E-04	-1.215483899	\\
1.29E-04	-1.177096792	\\
1.35E-04	-1.138709731	\\
1.42E-04	-1.094838758	\\
1.47E-04	-1.056451651	\\
1.54E-04	-1.015322634	\\
1.60E-04	-0.97419357	\\
2.32E-04	-0.933064553	\\
3.14E-04	-0.894677446	\\
4.02E-04	-0.853548476	\\
4.90E-04	-0.812419412	\\
5.70E-04	-0.774032305	\\
6.68E-04	-0.732903241	\\
7.61E-04	-0.691774271	\\
8.19E-04	-0.650645207	\\
8.03E-04	-0.6122581	\\
9.15E-04	-0.57112913	\\
8.97E-04	-0.530000066	\\
9.81E-04	-0.491612959	\\
1.04E-03	-0.450483989	\\
1.09E-03	-0.409354925	\\
1.14E-03	-0.370967818	\\
1.19E-03	-0.329838754	\\
1.35E-03	-0.288709784	\\
1.42E-03	-0.24758072	\\
1.55E-03	-0.209193613	\\
1.81E-03	-0.168064643	\\
2.25E-03	-0.126935579	\\
5.36E-03	-0.047419408	\\
9.58E-03	0.034838532	\\
1.41E-02	0.114354703	\\
1.91E-02	0.193870874	\\
2.40E-02	0.276129002	\\
2.77E-02	0.355644985	\\
2.99E-02	0.437903113	\\
3.04E-02	0.517419284	\\
2.93E-02	0.596935268	\\
2.65E-02	0.679193395	\\
2.20E-02	0.758709566	\\
1.58E-02	0.840967506	\\
8.18E-03	0.920483677	\\
1.56E-08	0.999999849	\\
};
\addlegendentry{G=4}

\addplot [smooth, color=red, mark=*, mark size=1.5pt, mark options={solid, red}]
  table[row sep=crcr]{%
6.80E-07	-1.881355932	\\
6.70E-05	-1.794915262	\\
6.47E-05	-1.75457628	\\
3.79E-05	-1.751694927	\\
7.51E-05	-1.711355945	\\
1.39E-04	-1.668135597	\\
1.45E-04	-1.624915275	\\
3.34E-04	-1.58457628	\\
3.67E-04	-1.541355958	\\
4.13E-04	-1.49813561	\\
4.58E-04	-1.454915263	\\
5.01E-04	-1.414576318	\\
5.46E-04	-1.371355946	\\
5.91E-04	-1.328135623	\\
6.34E-04	-1.287796629	\\
6.91E-04	-1.247457684	\\
7.25E-04	-1.201355983	\\
7.55E-04	-1.161016989	\\
7.87E-04	-1.117796666	\\
8.20E-04	-1.074576294	\\
8.52E-04	-1.031355971	\\
8.82E-04	-0.991016977	\\
9.92E-04	-0.947796704	\\
1.03E-03	-0.904576331	\\
1.07E-03	-0.864237337	\\
1.12E-03	-0.821016965	\\
1.17E-03	-0.777796692	\\
1.21E-03	-0.73457632	\\
1.26E-03	-0.694237325	\\
1.30E-03	-0.651017052	\\
1.37E-03	-0.60779668	\\
1.44E-03	-0.567457685	\\
1.52E-03	-0.524237412	\\
1.60E-03	-0.48101704	\\
1.67E-03	-0.440678046	\\
1.74E-03	-0.397457674	\\
1.88E-03	-0.3542374	\\
2.30E-03	-0.311017028	\\
2.67E-03	-0.270678034	\\
3.06E-03	-0.22745776	\\
3.92E-03	-0.184237388	\\
6.55E-03	-0.100678022	\\
1.06E-02	-0.014237475	\\
1.64E-02	0.069321891	\\
2.33E-02	0.152881258	\\
2.99E-02	0.239322002	\\
3.48E-02	0.322881171	\\
3.79E-02	0.409321915	\\
3.88E-02	0.492881282	\\
3.75E-02	0.576440451	\\
3.39E-02	0.662881195	\\
2.82E-02	0.746440561	\\
2.02E-02	0.832881108	\\
1.05E-02	0.916440475	\\
2.06E-08	0.999999841	\\
};
\addlegendentry{G=5}

\end{axis}

\end{tikzpicture}

}
\label{fig:Interface-svf058}
}
\caption{(a) The normalised mobile layer thickness $h_{m} / h_{f}$ against the normalised fluid flux $q_{f} / q^{*}$ and (b) the normalised mobile layer thickness $h_{m} / d_{p}$ against the Shields number $\theta$. Hollow markers: the experimental data by Aussillous et al. \cite{Aussillous2013}. Filled circles: the DNS results from Kidanemariam \& Uhlmann \cite{Kidanemariam2014b}. The streamwise velocity profiles of the sediment cross-sections at varying fluid flow rates when $\phi = 0.45$, $0.52$, and $0.58$ are shown in (c), (d), and (e), respectively.}
\label{fig:Sediment-Results}
\end{figure}

Sediments with three solid volume fractions $\phi = 0.45, 0.52$ and $0.58$ were simulated and their mobilities are plotted in Figs. \ref{fig:Sediment-hm-q} and \ref{fig:Sediment-hm-theta}, with the experimental data of Aussillous et al. \cite{Aussillous2013} and the DNS results from Kidanemariam \& Uhlmann \cite{Kidanemariam2014b} incorporated for comparison. As expected, the thickness of the mobile layer increases monotonically with the fluid flux and with the Shields number which is defined as,

\begin{equation}
\theta = 6 \left( Re / Ga \right) \left( d_{p} / h_{f} \right)^{2}, 
\end{equation}

and decreases with $\phi$ under the same fluid flux. Nonlinear curves are obtained for the three solid volume fractions investigated in the present study. Good agreements with both experimental and DNS results are obtained by $\phi = 0.45$ and $\phi = 0.52$. Note that $\phi$ ranged between $0.43$ and $0.48$ in the simulations by Kidanemariam \& Uhlmann \cite{Kidanemariam2014b}, while it was assumed that $\phi = 0.585$ in the experiments by Aussillous et al. \cite{Aussillous2013} without precise measurements. However, the reasonable coverage of the experimental data over the results of $\phi = 0.45$ and $\phi = 0.52$ as well as the numerical results indicates that the actual $\phi$ in the experiments was likely to be smaller than the assumed $0.585$. The difference in $\phi$ leads to an obvious separation of the three curves in the present study as well as the deviation of $\phi = 0.58$ from the experimental data. The particle size dispersion might be another reason for the deviation of $\phi = 0.58$. Mono-sized particles were used by Aussillous et al. and Kidanemariam \& Uhlmann, while the particle size distribution follows $100$ to $70$ mesh sand in the present study to gain more practicability. The smaller particle fills in the voids of the sediment, which adds to the resistance of the sediment against shear.

The significant impact of solid volume fraction $\phi$ on sediment movement is also demonstrated by the streamwise velocity profile of sediment cross-section in Figs. \ref{fig:Interface-svf045} to \ref{fig:Interface-svf058}, in which the $y$-coordinate is normalised so that the fluid-sediment interface is located at $\left( y - h_{s} \right) / h_{f} = 0$, where $h_{s}$ and $h_{f}$ are the final thickness of the sediment and the final height of the pure fluid region, respectively. When the same body force is applied, sediment with smaller $\phi$ suffers from severer erosion, which results in greater mixture velocity and thus larger fluid flux.

An outstanding merit of the MPSM is the accurate evaluation of $\phi$ within the sediment. The mean solid volume fraction at position $h = y$ can be straightforwardly obtained by simply summing $\gamma$ of the LBM computational cells across the $xz$-plane and divide by the total number of nodes on the plane,

\begin{equation}
\phi \left( y \right) = \frac{1}{N_{x} N_{z}}\sum_{i=1}^{N_{x}} \sum_{k=1}^{N_{z}} \gamma_{ik} \left( y \right),
\label{eq-phi}
\end{equation}

where $N_{x}$ and $N_{z}$ are the number of nodes along the channel length and width, respectively. Fig. \ref{fig:Sediment-svf} presents the variations of $\phi$ along the $y$-axis under different fluid flow rates. At $Re = 2.16$, sediments of $\phi = 0.45$ and $\phi = 0.52$ already entered the bedload transport regime, while the sediment of $\phi = 0.58$ was experiencing the incipient motion of surface particles. An interesting finding is that slight dilation is exhibited by the case $\phi = 0.58$ as the position of the fluid-sediment interface moves above the original sediment height. At a high packing fraction, such dilatancy can cause shear-thickening of the granular material (e.g. see \cite{Brown2012} and \cite{Morris2009}), which inspires the idea of incorporating the non-Newtonian behaviour to sediment transport modelling. At a larger $Re = 8.07$, all three sediments are much eroded and $\phi$ starts to decrease at a lower height. When in motion, $\phi$ remains nearly constant from the bottom to approximately $2 \sim 4$ particle diameters away from the sediment surface where $\phi$ vanishes, which is consistent with the experimental observations by Aussillous et al. \cite{Aussillous2013}.

\begin{figure}
\centering
{
\resizebox{0.4\textwidth}{!}{%
\begin{tikzpicture}

\begin{axis}[%
width=0.45\textwidth,
height=0.35\textwidth,
scale only axis,
xmin=0,
xmax=0.8,
ymin=0,
ymax=20,
xtick={0,0.2,0.4,0.6,0.8},
ytick={0,5,10,15,20},
xmajorgrids=true,
xminorgrids=true,
ymajorgrids=true,
grid style=dashed,
xlabel={$\phi$},
label style={font=\large},
ylabel={$y / d_{p}$},
label style={font=\large},
axis background/.style={fill=white},
legend style={legend pos=north east, legend cell align=left, align=right, draw=black}
]

\addplot [smooth, color=black!30!green, mark=triangle, mark size=1.5pt, mark options={solid, black!30!green}]
  table[row sep=crcr]{%
0.190740467	0.10989011	\\
0.439556016	0.549450549	\\
0.4632198	1.098901099	\\
0.47724628	1.648351648	\\
0.4732232	2.197802198	\\
0.481456743	2.747252747	\\
0.4832266	3.296703297	\\
0.485591514	3.846153846	\\
0.47323	4.395604396	\\
0.47282991	4.945054945	\\
0.4832334	5.494505495	\\
0.470228561	6.043956044	\\
0.4732368	6.593406593	\\
0.469586885	7.142857143	\\
0.4632402	7.692307692	\\
0.466948098	8.241758242	\\
0.4532436	8.791208791	\\
0.447705695	9.340659341	\\
0.418524205	9.89010989	\\
0.389799294	10.16483516	\\
0.291003156	10.71428571	\\
0.043400748	11.26373626	\\
0.000326053	11.53846154	\\
0	12.08791209	\\
};
\addlegendentry{$\phi=0.45$}

\addplot [smooth, color=blue, mark=square*, mark size=1.5pt, mark options={solid, blue}]
  table[row sep=crcr]{%
0.246003207	0.10989011	\\
0.50167069	0.549450549	\\
0.513247	1.098901099	\\
0.524985473	1.648351648	\\
0.5232504	2.197802198	\\
0.525885261	2.747252747	\\
0.5332538	3.296703297	\\
0.539817085	3.846153846	\\
0.5332572	4.395604396	\\
0.52559808	4.945054945	\\
0.5232606	5.494505495	\\
0.524584342	6.043956044	\\
0.533264	6.593406593	\\
0.523823949	7.142857143	\\
0.5132674	7.692307692	\\
0.511858036	8.241758242	\\
0.5032708	8.791208791	\\
0.48562501	9.340659341	\\
0.458870977	10.16483516	\\
0.328108931	10.71428571	\\
0.222997548	11.26373626	\\
0.116178148	11.81318681	\\
0.000413816	12.36263736	\\
0	12.63736264	\\
};
\addlegendentry{$\phi=0.52$}

\addplot [smooth, color=red, mark=*, mark size=1.5pt, mark options={solid, red}]
  table[row sep=crcr]{%
0.325716484	0.10989011	\\
0.561423248	0.549450549	\\
0.5832742	1.098901099	\\
0.594605852	1.648351648	\\
0.5932776	2.197802198	\\
0.59083674	2.747252747	\\
0.593281	3.296703297	\\
0.590826232	3.846153846	\\
0.5832844	4.395604396	\\
0.577644123	4.945054945	\\
0.5732878	5.494505495	\\
0.575586754	6.043956044	\\
0.5732912	6.593406593	\\
0.577345943	7.142857143	\\
0.5632946	7.692307692	\\
0.562194309	8.241758242	\\
0.553298	8.791208791	\\
0.549885555	9.340659341	\\
0.536500673	10.16483516	\\
0.429557372	10.71428571	\\
0.25553785	11.26373626	\\
0.085120666	11.81318681	\\
0.001808	12.36263736	\\
0	12.63736264	\\
};
\addlegendentry{$\phi=0.58$}

\addplot [smooth, black, dashed, line width = 1.5]
  table[row sep=crcr]{%
0	10.989 \\
0.8	10.989 \\
};

\end{axis}

\end{tikzpicture}

}
\label{fig:svf-vs-y-g1}
}
{
\resizebox{0.4\textwidth}{!}{%
\begin{tikzpicture}

\begin{axis}[%
width=0.45\textwidth,
height=0.35\textwidth,
scale only axis,
xmin=0,
xmax=0.8,
ymin=0,
ymax=20,
xtick={0,0.2,0.4,0.6,0.8},
ytick={0,5,10,15,20},
xmajorgrids=true,
xminorgrids=true,
ymajorgrids=true,
grid style=dashed,
xlabel={$\phi$},
label style={font=\large},
ylabel={$y / d_{p}$},
label style={font=\large},
axis background/.style={fill=white},
legend style={legend pos=north east, legend cell align=left, align=right, draw=black}
]

\addplot [smooth, color=black!30!green, mark=triangle, mark size=1.5pt, mark options={solid, black!30!green}]
  table[row sep=crcr]{%
0.176889654	0.10989011	\\
0.469321522	0.549450549	\\
0.481482906	1.098901099	\\
0.467246283	1.648351648	\\
0.471441906	2.197802198	\\
0.481456746	2.747252747	\\
0.471399955	3.296703297	\\
0.485297445	3.846153846	\\
0.481358004	4.395604396	\\
0.472829919	4.945054945	\\
0.471316053	5.494505495	\\
0.470228564	6.043956044	\\
0.461232153	6.593406593	\\
0.469586889	7.142857143	\\
0.461274104	7.692307692	\\
0.461700829	8.241758242	\\
0.45118965	8.791208791	\\
0.447705706	9.340659341	\\
0.41108761	9.89010989	\\
0.388230615	10.16483516	\\
0.29100317	10.71428571	\\
0.041804621	11.26373626	\\
0.000326053	11.53846154	\\
0	12.08791209	\\
};
\addlegendentry{$\phi=0.45$}

\addplot [smooth, color=blue, mark=square*, mark size=1.5pt, mark options={solid, blue}]
  table[row sep=crcr]{%
0.225629885	0.10989011	\\
0.531895527	0.549450549	\\
0.521146357	1.098901099	\\
0.52152325	1.648351648	\\
0.531103065	2.197802198	\\
0.525737493	2.747252747	\\
0.531059772	3.296703297	\\
0.524247911	3.846153846	\\
0.521016479	4.395604396	\\
0.527006889	4.945054945	\\
0.520973187	5.494505495	\\
0.514006546	6.043956044	\\
0.522929181	6.593406593	\\
0.50937278	7.142857143	\\
0.510884545	7.692307692	\\
0.498847026	8.241758242	\\
0.480839905	8.791208791	\\
0.462945013	9.340659341	\\
0.412037484	9.89010989	\\
0.360247905	10.16483516	\\
0.306053739	10.71428571	\\
0.237900499	11.26373626	\\
0.169330958	11.81318681	\\
0.059115488	12.36263736	\\
0.019640117	12.91208791	\\
0.009390232	13.18681319	\\
0	13.46153846	\\
};
\addlegendentry{$\phi=0.52$}

\addplot [smooth, color=red, mark=*, mark size=1.5pt, mark options={solid, red}]
  table[row sep=crcr]{%
0.325716484	0.10989011	\\
0.591231955	0.549450549	\\
0.580795266	1.098901099	\\
0.594605851	1.648351648	\\
0.587506265	2.197802198	\\
0.590836743	2.747252747	\\
0.590705987	3.296703297	\\
0.591143458	3.846153846	\\
0.590660479	4.395604396	\\
0.577644136	4.945054945	\\
0.580614497	5.494505495	\\
0.575586762	6.043956044	\\
0.580685145	6.593406593	\\
0.577345947	7.142857143	\\
0.570522532	7.692307692	\\
0.562194312	8.241758242	\\
0.560476553	8.791208791	\\
0.531885556	9.340659341	\\
0.503553915	9.89010989	\\
0.471601464	10.16483516	\\
0.411717372	10.71428571	\\
0.228541755	11.26373626	\\
0.110306597	11.81318681	\\
0.01241259	12.36263736	\\
0.002195695	12.91208791	\\
0	13.18681319	\\
};
\addlegendentry{$\phi=0.58$}

\addplot [smooth, black, dashed, line width = 1.5]
  table[row sep=crcr]{%
0	10.989 \\
0.8	10.989 \\
};

\end{axis}

\end{tikzpicture}

}
\label{fig:svf-vs-y-g3}
}
\caption{The variations of solid volume fraction $\phi$ along the $y$-axis for (left) $Re = 2.16$ and (right) $Re = 8.07$. The dashed line ($y / d_{p} = 11$) represent the initial fluid-sediment interface prior to shearing.}
\label{fig:Sediment-svf}
\end{figure}
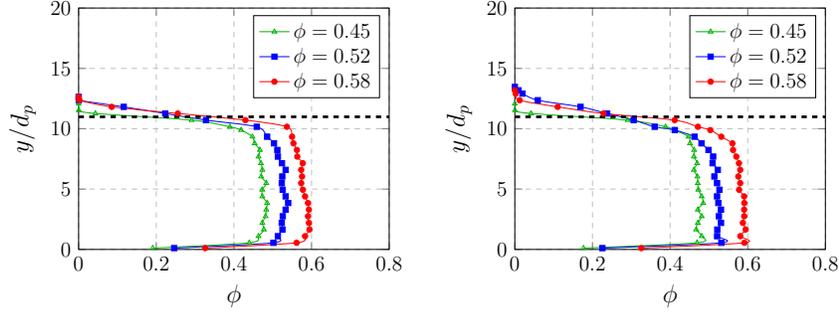

\subsection{The particle flux}
\label{sec:sediment-qp}

\begin{figure}
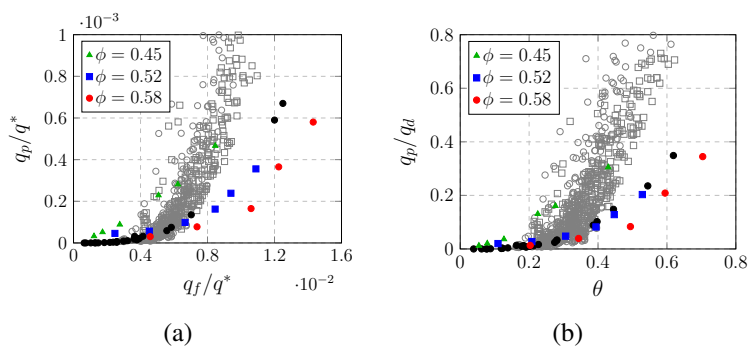

\centering
\subfigure[]{
\resizebox{0.35\textwidth}{!}{%
\input{./Figures/Sediment-qp-q.tex}}
\label{fig:Sediment-qp-q}
}
\subfigure[]{
\resizebox{0.36\textwidth}{!}{%
\input{./Figures/Sediment-qp-theta.tex}}
\label{fig:Sediment-qp-theta}
}
\caption{Results of the sediment movement modelling, showing (a) the normalised particle flux $q_{p} / q^{*}$ against the normalised fluid flux $q_{f} / q^{*}$ and (b) the normalised particle flux $q_{p} / q_{d}$ against the Shields number $\theta$. The hollow markers correspond to the experimental data reported by Aussillous et al. \cite{Aussillous2013} and the filled circles represent the DNS results from Kidanemariam \& Uhlmann \cite{Kidanemariam2014b}.}
\label{fig:Sediment-qp}
\end{figure}

The volumetric particle flow rate, $q_{p}$, is calculated by \cite{Kidanemariam2014b},

\begin{equation}
q_{p} = \frac{\pi}{6 L_{x} L_{z}} \sum u_{p} d_{p}^{3},
\end{equation}

\noindent where $u_{p}$ is the streamwise velocity of the particle and $L_{x}$ and $L_{z}$ are with respect to the length and width of the channel. The results for particle flux are plotted in Fig. \ref{fig:Sediment-qp}. Again a near-linear curve was obtained from $\phi = 0.45$ and quadratic curves were formulated by $\phi = 0.52$ and $\phi = 0.58$, which is consistent with the observations in S~\ref{sec:sediment-hm}.

An outstanding merit of the MPSM is the straightforward, accurate evaluation of $\phi$ within the sediment. The mean solid volume fraction at position $h = y$ was calculated by summing $\gamma$ of the LBM computational cells across the $xz$-plane once the fluid and particle fluxes reached a steady state, i.e.

\begin{equation}
\phi \left( y \right) = \frac{1}{N_{x} N_{z}}\sum_{i=1}^{N_{x}} \sum_{k=1}^{N_{z}} \gamma_{ik} \left( y \right),
\end{equation}

\noindent where $N_{x}$ and $N_{z}$ are the number of nodes along the channel length and width, respectively. We present the variations of $\phi$ along $y$-axis under different fluid flow rates in Fig. \ref{fig:Sediment-svf}. At $Re = 2.16$, sediments of $\phi = 0.45$ and $\phi = 0.52$ entered the 'flat bed motion', while the sediment of $\phi = 0.58$ was experiencing the incipient motion of the particles. An interesting finding was that slight dilatancy was exhibited by the case $\phi = 0.58$ as the position of the fluid-sediment interface moved above the original sediment height. In rheological studies, such dilatancy indicated shear-thickening of the material, which motivated the idea of incorporating the non-Newtonian behaviour to sediment movement modelling. At a larger $Re = 8.07$, all three sediments were much eroded and the computed fluid-sediment interface moved downwards. When in motion, $\phi$ remained nearly constant from the bottom to approximately two particle diameters away from the sediment surface, which is consistent with the experimental observations by Aussillous et al. \cite{Aussillous2013}.

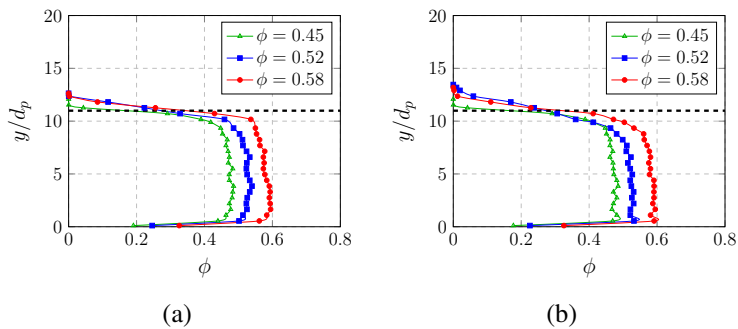
\begin{figure}
\centering
\subfigure[]{
\resizebox{0.35\textwidth}{!}{%
\begin{tikzpicture}

\begin{axis}[%
width=0.45\textwidth,
height=0.35\textwidth,
scale only axis,
xmin=0,
xmax=0.8,
ymin=0,
ymax=20,
xtick={0,0.2,0.4,0.6,0.8},
ytick={0,5,10,15,20},
xmajorgrids=true,
xminorgrids=true,
ymajorgrids=true,
grid style=dashed,
xlabel={$\phi$},
label style={font=\large},
ylabel={$y / d_{p}$},
label style={font=\large},
axis background/.style={fill=white},
legend style={legend pos=north east, legend cell align=left, align=right, draw=black}
]

\addplot [smooth, color=black!30!green, mark=triangle, mark size=1.5pt, mark options={solid, black!30!green}]
  table[row sep=crcr]{%
0.190740467	0.10989011	\\
0.439556016	0.549450549	\\
0.4632198	1.098901099	\\
0.47724628	1.648351648	\\
0.4732232	2.197802198	\\
0.481456743	2.747252747	\\
0.4832266	3.296703297	\\
0.485591514	3.846153846	\\
0.47323	4.395604396	\\
0.47282991	4.945054945	\\
0.4832334	5.494505495	\\
0.470228561	6.043956044	\\
0.4732368	6.593406593	\\
0.469586885	7.142857143	\\
0.4632402	7.692307692	\\
0.466948098	8.241758242	\\
0.4532436	8.791208791	\\
0.447705695	9.340659341	\\
0.418524205	9.89010989	\\
0.389799294	10.16483516	\\
0.291003156	10.71428571	\\
0.043400748	11.26373626	\\
0.000326053	11.53846154	\\
0	12.08791209	\\
};
\addlegendentry{$\phi=0.45$}

\addplot [smooth, color=blue, mark=square*, mark size=1.5pt, mark options={solid, blue}]
  table[row sep=crcr]{%
0.246003207	0.10989011	\\
0.50167069	0.549450549	\\
0.513247	1.098901099	\\
0.524985473	1.648351648	\\
0.5232504	2.197802198	\\
0.525885261	2.747252747	\\
0.5332538	3.296703297	\\
0.539817085	3.846153846	\\
0.5332572	4.395604396	\\
0.52559808	4.945054945	\\
0.5232606	5.494505495	\\
0.524584342	6.043956044	\\
0.533264	6.593406593	\\
0.523823949	7.142857143	\\
0.5132674	7.692307692	\\
0.511858036	8.241758242	\\
0.5032708	8.791208791	\\
0.48562501	9.340659341	\\
0.458870977	10.16483516	\\
0.328108931	10.71428571	\\
0.222997548	11.26373626	\\
0.116178148	11.81318681	\\
0.000413816	12.36263736	\\
0	12.63736264	\\
};
\addlegendentry{$\phi=0.52$}

\addplot [smooth, color=red, mark=*, mark size=1.5pt, mark options={solid, red}]
  table[row sep=crcr]{%
0.325716484	0.10989011	\\
0.561423248	0.549450549	\\
0.5832742	1.098901099	\\
0.594605852	1.648351648	\\
0.5932776	2.197802198	\\
0.59083674	2.747252747	\\
0.593281	3.296703297	\\
0.590826232	3.846153846	\\
0.5832844	4.395604396	\\
0.577644123	4.945054945	\\
0.5732878	5.494505495	\\
0.575586754	6.043956044	\\
0.5732912	6.593406593	\\
0.577345943	7.142857143	\\
0.5632946	7.692307692	\\
0.562194309	8.241758242	\\
0.553298	8.791208791	\\
0.549885555	9.340659341	\\
0.536500673	10.16483516	\\
0.429557372	10.71428571	\\
0.25553785	11.26373626	\\
0.085120666	11.81318681	\\
0.001808	12.36263736	\\
0	12.63736264	\\
};
\addlegendentry{$\phi=0.58$}

\addplot [smooth, black, dashed, line width = 1.5]
  table[row sep=crcr]{%
0	10.989 \\
0.8	10.989 \\
};

\end{axis}

\end{tikzpicture}

}
\label{fig:svf-vs-y-g1}
}
\subfigure[]{
\resizebox{0.35\textwidth}{!}{%
\begin{tikzpicture}

\begin{axis}[%
width=0.45\textwidth,
height=0.35\textwidth,
scale only axis,
xmin=0,
xmax=0.8,
ymin=0,
ymax=20,
xtick={0,0.2,0.4,0.6,0.8},
ytick={0,5,10,15,20},
xmajorgrids=true,
xminorgrids=true,
ymajorgrids=true,
grid style=dashed,
xlabel={$\phi$},
label style={font=\large},
ylabel={$y / d_{p}$},
label style={font=\large},
axis background/.style={fill=white},
legend style={legend pos=north east, legend cell align=left, align=right, draw=black}
]

\addplot [smooth, color=black!30!green, mark=triangle, mark size=1.5pt, mark options={solid, black!30!green}]
  table[row sep=crcr]{%
0.176889654	0.10989011	\\
0.469321522	0.549450549	\\
0.481482906	1.098901099	\\
0.467246283	1.648351648	\\
0.471441906	2.197802198	\\
0.481456746	2.747252747	\\
0.471399955	3.296703297	\\
0.485297445	3.846153846	\\
0.481358004	4.395604396	\\
0.472829919	4.945054945	\\
0.471316053	5.494505495	\\
0.470228564	6.043956044	\\
0.461232153	6.593406593	\\
0.469586889	7.142857143	\\
0.461274104	7.692307692	\\
0.461700829	8.241758242	\\
0.45118965	8.791208791	\\
0.447705706	9.340659341	\\
0.41108761	9.89010989	\\
0.388230615	10.16483516	\\
0.29100317	10.71428571	\\
0.041804621	11.26373626	\\
0.000326053	11.53846154	\\
0	12.08791209	\\
};
\addlegendentry{$\phi=0.45$}

\addplot [smooth, color=blue, mark=square*, mark size=1.5pt, mark options={solid, blue}]
  table[row sep=crcr]{%
0.225629885	0.10989011	\\
0.531895527	0.549450549	\\
0.521146357	1.098901099	\\
0.52152325	1.648351648	\\
0.531103065	2.197802198	\\
0.525737493	2.747252747	\\
0.531059772	3.296703297	\\
0.524247911	3.846153846	\\
0.521016479	4.395604396	\\
0.527006889	4.945054945	\\
0.520973187	5.494505495	\\
0.514006546	6.043956044	\\
0.522929181	6.593406593	\\
0.50937278	7.142857143	\\
0.510884545	7.692307692	\\
0.498847026	8.241758242	\\
0.480839905	8.791208791	\\
0.462945013	9.340659341	\\
0.412037484	9.89010989	\\
0.360247905	10.16483516	\\
0.306053739	10.71428571	\\
0.237900499	11.26373626	\\
0.169330958	11.81318681	\\
0.059115488	12.36263736	\\
0.019640117	12.91208791	\\
0.009390232	13.18681319	\\
0	13.46153846	\\
};
\addlegendentry{$\phi=0.52$}

\addplot [smooth, color=red, mark=*, mark size=1.5pt, mark options={solid, red}]
  table[row sep=crcr]{%
0.325716484	0.10989011	\\
0.591231955	0.549450549	\\
0.580795266	1.098901099	\\
0.594605851	1.648351648	\\
0.587506265	2.197802198	\\
0.590836743	2.747252747	\\
0.590705987	3.296703297	\\
0.591143458	3.846153846	\\
0.590660479	4.395604396	\\
0.577644136	4.945054945	\\
0.580614497	5.494505495	\\
0.575586762	6.043956044	\\
0.580685145	6.593406593	\\
0.577345947	7.142857143	\\
0.570522532	7.692307692	\\
0.562194312	8.241758242	\\
0.560476553	8.791208791	\\
0.531885556	9.340659341	\\
0.503553915	9.89010989	\\
0.471601464	10.16483516	\\
0.411717372	10.71428571	\\
0.228541755	11.26373626	\\
0.110306597	11.81318681	\\
0.01241259	12.36263736	\\
0.002195695	12.91208791	\\
0	13.18681319	\\
};
\addlegendentry{$\phi=0.58$}

\addplot [smooth, black, dashed, line width = 1.5]
  table[row sep=crcr]{%
0	10.989 \\
0.8	10.989 \\
};

\end{axis}

\end{tikzpicture}

}
\label{fig:svf-vs-y-g3}
}
\caption{The variations of $\phi$ along $y$-axis for (a) $Re = 2.16$ and (b) $Re = 8.07$.}
\label{fig:Sediment-svf}
\end{figure}

\subsection{The effective viscosity and friction coefficient}
\label{sec:sediment-nu}

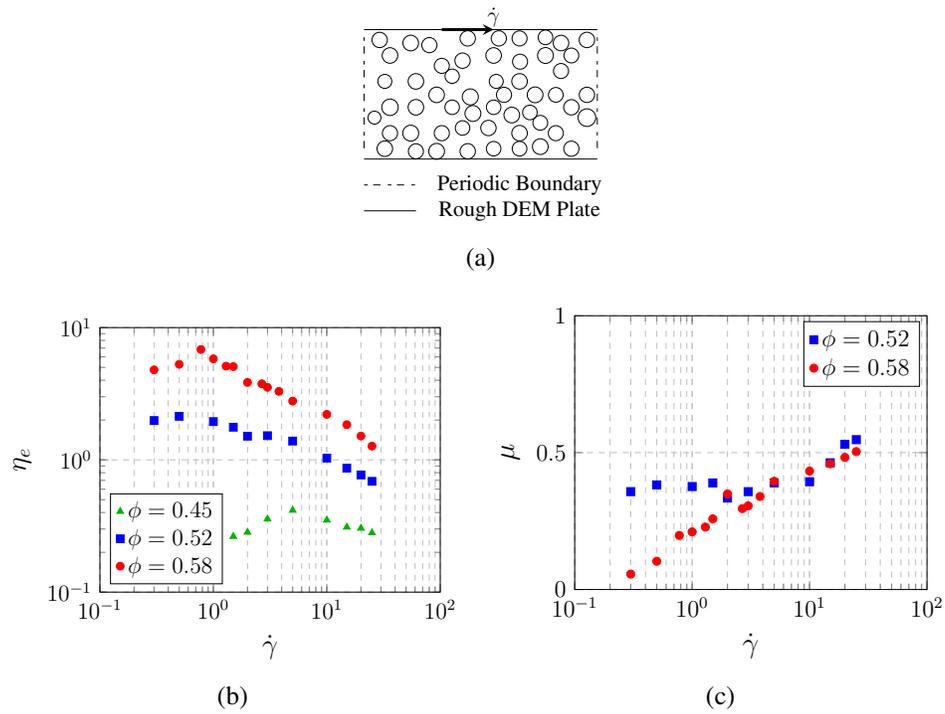
\begin{figure}
\centering
\subfigure[]{
\resizebox{0.25\textwidth}{!}{%
\begin{tikzpicture}

\draw (-2.5,3.5) node (v1) {} -- (2,3.5) node (v3) {};
\draw (-2.5,1) node (v2) {} -- (2,1) node (v4) {};


\draw  (0,2) ellipse (0.14 and 0.14);
\draw  (1.8,1.8) ellipse (0.17 and 0.17);
\draw  (-2.3,1.8) ellipse (0.125 and 0.125);
\draw  (-2.1,2.5) ellipse (0.135 and 0.135);
\draw  (0.05,2.5) ellipse (0.135 and 0.135);
\draw  (-1,2.8) ellipse (0.145 and 0.145);
\draw  (-0.8,2.6) ellipse (0.138 and 0.138);
\draw  (-0.6,2.9) ellipse (0.148 and 0.148);
\draw  (1,3) ellipse (0.153 and 0.153);
\draw  (-2,2) ellipse (0.159 and 0.159);
\draw  (-1,1.5) ellipse (0.146 and 0.146);
\draw  (1.5,2) ellipse (0.148 and 0.148);
\draw  (-2,1.5) ellipse (0.146 and 0.146);
\draw  (-1.5,2.5) ellipse (0.15 and 0.15);
\draw  (-1.5,2) ellipse (0.15 and 0.15);
\draw  (0,3) ellipse (0.15 and 0.15);
\draw  (0.5,2.5) ellipse (0.15 and 0.15);
\draw  (0.5,1.5) ellipse (0.15 and 0.15);
\draw  (-2,3) ellipse (0.15 and 0.15);
\draw  (-1.24,3.2) ellipse (0.146 and 0.146);
\draw  (-0.8,2) ellipse (0.14 and 0.14);
\draw  (-0.6,1.6) ellipse (0.14 and 0.14);
\draw  (-0.1,1.6) ellipse (0.148 and 0.148);
\draw  (1.2,2.24) ellipse (0.142 and 0.142);
\draw  (0.2,2.24) ellipse (0.146 and 0.146);
\draw  (1.8,2.24) ellipse (0.148 and 0.148);
\draw  (-1.1,2.24) ellipse (0.15 and 0.15);
\draw  (0.8,2.24) ellipse (0.152 and 0.152);
\draw  (0.9,1.24) ellipse (0.152 and 0.152);
\draw  (-1.1,1.15) ellipse (0.152 and 0.152);
\draw  (-1.5,1.15) ellipse (0.145 and 0.145);
\draw  (-2.1,1.2) ellipse (0.15 and 0.15);
\draw  (-0.5,1.15) ellipse (0.148 and 0.148);
\draw  (0,1.2) ellipse (0.14 and 0.14);
\draw  (0.5,1.2) ellipse (0.142 and 0.142);
\draw  (1.5,1.2) ellipse (0.147 and 0.147);
\draw  (1.2,1.5) ellipse (0.145 and 0.145);
\draw  (0.9,1.7) ellipse (0.143 and 0.143);
\draw  (0.7,1.9) ellipse (0.141 and 0.141);
\draw  (0.35,1.85) ellipse (0.151 and 0.151);
\draw  (-0.4,1.88) ellipse (0.153 and 0.153);
\draw  (-0.45,2.2) ellipse (0.155 and 0.155);
\draw  (-1.6,1.5) ellipse (0.154 and 0.154);
\draw  (1.3,3.24) ellipse (0.15 and 0.15);
\draw  (-1.6,3.24) ellipse (0.151 and 0.151);
\draw  (-2.2,3.3) ellipse (0.148 and 0.148);
\draw  (1.8,3.3) ellipse (0.149 and 0.149);
\draw  (1.5,3) ellipse (0.143 and 0.143);
\draw  (1.3,2.7) ellipse (0.145 and 0.145);
\draw  (1,3.35) ellipse (0.141 and 0.141);
\draw  (0.5,3.33) ellipse (0.145 and 0.145);
\draw  (0.1,3.33) ellipse (0.147 and 0.147);
\draw  (-0.5,3.33) ellipse (0.148 and 0.148);
\draw  (0.51,2.88) ellipse (0.145 and 0.145);

\draw [dash pattern=on 2pt off 3pt on 4pt off 4pt](v1) -- (v2);
\draw [dash pattern=on 2pt off 3pt on 4pt off 4pt](v3) -- (v4);

\draw [ultra thick,>=stealth, ->] (-1,3.5) -- (0,3.5);
\node at (0,3.75) {$\dot{\gamma}$};


\draw [dash pattern=on 2pt off 3pt on 4pt off 4pt](-2.5,0.5) -- (-1.5,0.5);
\node at (0.5,0.5) {Periodic Boundary};

\draw (-2.5,0) -- (-1.5,0);
\node at (0.5,0) {Rough DEM Plate};

\end{tikzpicture}
}
\label{fig:rheometer}
}\\
\subfigure[]{
\resizebox{0.45\textwidth}{!}{%
\begin{tikzpicture}

\begin{axis}[%
width=0.45\textwidth,
height=0.35\textwidth,
scale only axis,
xmode=log,
ymode=log,
xmin=0.1,
xmax=100,
ymin=0.1,
ymax=10,
xtick={0.1,1,10,100},
ytick={0.1,1,10},
xmajorgrids=true,
xminorgrids=true,
ymajorgrids=true,
grid style=dashed,
xlabel={$\dot{\gamma}$},
label style={font=\large},
ylabel={$\eta_{e}$},
label style={font=\large},
axis background/.style={fill=white},
legend style={legend pos=south west, legend cell align=left, align=right, draw=black}
]

\addplot [only marks, color=black!30!green, mark=triangle*, mark size=2pt, mark options={solid, black!30!green}]
  table[row sep=crcr]{%
0.3	0.134148057	\\
0.5	0.142028994	\\
1	0.194861163	\\
1.5	0.263532933	\\
2	0.28340233	\\
3	0.357095672	\\
5	0.415381331	\\
10	0.350438183	\\
15	0.309981665	\\
20	0.304017486	\\
25	0.280749941	\\
};
\addlegendentry{$\phi=0.45$}

\addplot [only marks, color=blue, mark=square*, mark size=2pt, mark options={solid, blue}]
  table[row sep=crcr]{%
0.3	1.983295865	\\
0.5	2.134139984	\\
1	1.94647547	\\
1.5	1.765591001	\\
2	1.510780222	\\
3	1.523595828	\\
5	1.387307717	\\
10	1.031488706	\\
15	0.86586399	\\
20	0.769262736	\\
25	0.688050393	\\
};
\addlegendentry{$\phi=0.52$}

\addplot [only marks, color=red, mark=*, mark size=2pt, mark options={solid, red}]
  table[row sep=crcr]{%
0.3	4.789503945	\\
0.5	5.275548562	\\
0.777202	6.828358374	\\
1	5.803366535	\\
1.295337	5.109609306	\\
1.5	5.06974114	\\
2	3.851417033	\\
2.673797	3.758014339	\\
3	3.53211024	\\
3.783784	3.29466164	\\
5	2.782856769	\\
10	2.207360849	\\
15	1.844984814	\\
20	1.513853935	\\
25	1.270581803	\\
};
\addlegendentry{$\phi=0.58$}

\end{axis}

\end{tikzpicture}

}
\label{fig:Sediment-viscosity}
}
\subfigure[]{
\resizebox{0.45\textwidth}{!}{%
\begin{tikzpicture}

\begin{axis}[%
width=0.45\textwidth,
height=0.35\textwidth,
scale only axis,
xmode=log,
xmin=0.1,
xmax=100,
ymin=0,
ymax=1,
xtick={0.1,1,10,100},
ytick={0,0.5,1},
xmajorgrids=true,
xminorgrids=true,
ymajorgrids=true,
grid style=dashed,
xlabel={$\dot{\gamma}$},
label style={font=\large},
ylabel={$\mu$},
label style={font=\large},
axis background/.style={fill=white},
]

\addplot [only marks, color=blue, mark=square*, mark size=2pt, mark options={solid, blue}]
  table[row sep=crcr]{%
0.3	0.357575993	\\
0.5	0.382173015	\\
1	0.376157969	\\
1.5	0.389412446	\\
2	0.334168959	\\
3	0.35755238	\\
5	0.389751752	\\
10	0.393790209	\\
15	0.462975145	\\
20	0.530286719	\\
25	0.547326217	\\
};
\addlegendentry{$\phi=0.52$}

\addplot [only marks, color=red, mark=*, mark size=2pt, mark options={solid, red}]
  table[row sep=crcr]{%
0.3	0.056323798	\\
0.5	0.103337454	\\
0.777202	0.19758206	\\
1	0.21052736	\\
1.295337	0.228085969	\\
1.5	0.258126359	\\
2	0.348925634	\\
2.673797	0.29548709	\\
3	0.305678986	\\
3.783784	0.339815807	\\
5	0.395210086	\\
10	0.43275365	\\
15	0.459343939	\\
20	0.482577845	\\
25	0.503902995	\\
};
\addlegendentry{$\phi=0.58$}

\end{axis}

\end{tikzpicture}

}
\label{fig:Sediment-mu}
}
\caption{(a) A section-view schematic diagram of the shear cell. Shear-rate-dependent properties of the sediment with varying $\phi$ are shown in (b) for the effective viscosity and (c) for the friction coefficient.}
\label{fig:Sediment-rheology}
\end{figure}

Motivated by the dilation observed for $\phi = 0.58$ at low Reynolds number, we further quantify the shear-rate-dependent effective viscosity and friction coefficient of the sediment via a shear cell test. As shown in Fig. \ref{fig:rheometer}, the test case featured a moving plate on top of the sediment, to which a constant shear rate $\dot{\gamma}$ was assigned according to the fluid-sediment interface velocity interpreted from the velocity profiles. The height of the shear cell varied in each case according to the final position of the fluid-sediment interface to include sediment dilation and solid volume fraction gradient near the interface. The normal particle pressure $p^{p}$ and the shear stress $\tau$ acting on the plate were recorded for computations which are given by,

\begin{equation}
   \begin{aligned}
   \eta_{e} = \tau / \dot{\gamma} &&\textup{and}&& \mu = \tau / p^{p},
   \end{aligned}
\label{eq-rheometer}
\end{equation}

\noindent for the effective viscosity $\eta_{e}$ and the friction coefficient $\mu$, respectively.

Our numerical results are summarised in Figs. \ref{fig:Sediment-viscosity} and \ref{fig:Sediment-mu}. After experiencing the initial shear-thickening due to confined dilatancy at small shear rates, a strong shear-thinning rheology is observed at larger shear rates. Meanwhile, the friction coefficient also exhibits shear-rate dependency. At small fluid flow rates, the sediment first exhibits shear-thickening rheology, due to which the thickness of the mobile layer slowly increases with the fluid flux. With the increment of the fluid flow rate, the fluid-sediment interface velocity increases, thus exerting a larger shear rate over the sediment. The sediment enters the shear-thinning rheology and the resistance to shearing flow becomes weaker. Meanwhile, the increase of the friction coefficient intensifies the shear stress which the upper fluid exerts on the sediment surface. Consequently, a dramatic increase in $h_{m}$ is observed with the amplification of $q_{f}$. 

\begin{figure}
\centering
\subfigure[]{
\resizebox{0.45\textwidth}{!}{%
\begin{tikzpicture}

\begin{axis}[%
width=0.45\textwidth,
height=0.35\textwidth,
scale only axis,
xmin=0,
xmax=0.01,
ymin=0,
ymax=1,
xtick={0,0.005,0.01},
ytick={0,0.5,1},
xmajorgrids=true,
xminorgrids=true,
ymajorgrids=true,
grid style=dashed,
xlabel={$u_{x} h_{f}/ q^{*}$},
label style={font=\large},
ylabel={$\left( y - h_{s} \right) / h_{f}$},
label style={font=\large},
axis background/.style={fill=white},
legend style={at={(0.005,0.5)}, anchor=west, legend cell align=left, align=right, draw=black}
]

\addplot [only marks, color=black!30!green, mark=triangle*, mark size=2pt, mark options={solid, black!30!green}]
  table[row sep=crcr]{%
2.77E-03	0.056376745	\\
4.16E-03	0.132753463	\\
5.19E-03	0.204202777	\\
5.97E-03	0.27565209	\\
6.51E-03	0.34956519	\\
6.78E-03	0.421014335	\\
6.80E-03	0.494927435	\\
6.57E-03	0.566376748	\\
6.08E-03	0.637825893	\\
5.32E-03	0.711738993	\\
4.33E-03	0.783188306	\\
3.05E-03	0.857101237	\\
1.55E-03	0.928550551	\\
3.14E-05	0.997536077	\\
};

\addplot [smooth, dashed, color=black!30!green, line width = 1.5pt]
  table[row sep=crcr]{%
1.51E-03	-0.012608781	\\  
2.99E-03	0.056376745	\\
4.35E-03	0.132753463	\\
5.37E-03	0.204202777	\\
6.13E-03	0.27565209	\\
6.66E-03	0.34956519	\\
6.91E-03	0.421014335	\\
6.90E-03	0.494927435	\\
6.63E-03	0.566376748	\\
6.12E-03	0.637825893	\\
5.32E-03	0.711738993	\\
4.29E-03	0.783188306	\\
2.96E-03	0.857101237	\\
1.41E-03	0.928550551	\\
-3.19E-04	0.997536077	\\
};

\addplot [only marks, color=blue, mark=square*, mark size=2pt, mark options={solid, blue}]
  table[row sep=crcr]{%
1.70E-03	0.044999952	\\
3.60E-03	0.119999838	\\
5.05E-03	0.192499876	\\
6.16E-03	0.264999915	\\
7.01E-03	0.339999972	\\
7.52E-03	0.41249984	\\
7.71E-03	0.487499897	\\
7.57E-03	0.559999936	\\
7.11E-03	0.632499803	\\
6.29E-03	0.70749986	\\
5.17E-03	0.779999899	\\
3.66E-03	0.854999785	\\
1.88E-03	0.927499823	\\
8.10E-09	0.999999862	\\
};

\addplot [smooth, dashed, color=blue, line width = 1.5pt]
  table[row sep=crcr]{%
3.955E-05	-0.027500087	\\
0.002144677	0.044999952	\\
0.003980322	0.119999838	\\
0.005424116	0.192499876	\\
0.006542848	0.264999915	\\
0.007358088	0.339999972	\\
0.007815486	0.41249984	\\
0.007946589	0.487499897	\\
0.007742654	0.559999936	\\
0.00721366	0.632499803	\\
0.006324353	0.70749986	\\
0.005134024	0.779999899	\\
0.003560584	0.854999785	\\
0.001708922	0.927499823	\\
-0.000467802	0.999999862	\\
};

\addplot [only marks, color=red, mark=*, mark size=2pt, mark options={solid, red}]
  table[row sep=crcr]{%
1.41E-03	0.058840532	\\
3.18E-03	0.132753463	\\
4.84E-03	0.204202777	\\
6.21E-03	0.27565209	\\
7.27E-03	0.34956519	\\
7.93E-03	0.421014335	\\
8.24E-03	0.494927435	\\
8.16E-03	0.566376748	\\
7.71E-03	0.637825893	\\
6.85E-03	0.711738993	\\
5.65E-03	0.783188306	\\
4.02E-03	0.857101237	\\
2.07E-03	0.928550551	\\
1.50E-08	0.999999864	\\
};

\addplot [smooth, dashed, color=red, line width = 1.5pt]
  table[row sep=crcr]{%
0.000242229	-0.012608781	\\
0.00247692	0.058840532	\\
0.004423217	0.132753463	\\
0.005951375	0.204202777	\\
0.007132254	0.27565209	\\
0.007988404	0.34956519	\\
0.008462749	0.421014335	\\
0.008588003	0.494927435	\\
0.008355816	0.566376748	\\
0.00777635	0.637825893	\\
0.006811454	0.711738993	\\
0.005525455	0.783188306	\\
0.003829665	0.857101237	\\
0.001837133	0.928550551	\\
-0.000502679	0.999999864	\\
};

\legend{$\phi=0.45$,,$\phi=0.52$,,$\phi=0.58$,};

\end{axis}

\end{tikzpicture}

}
\label{fig:purefluid}
}
\subfigure[]{
\resizebox{0.45\textwidth}{!}{%
\input{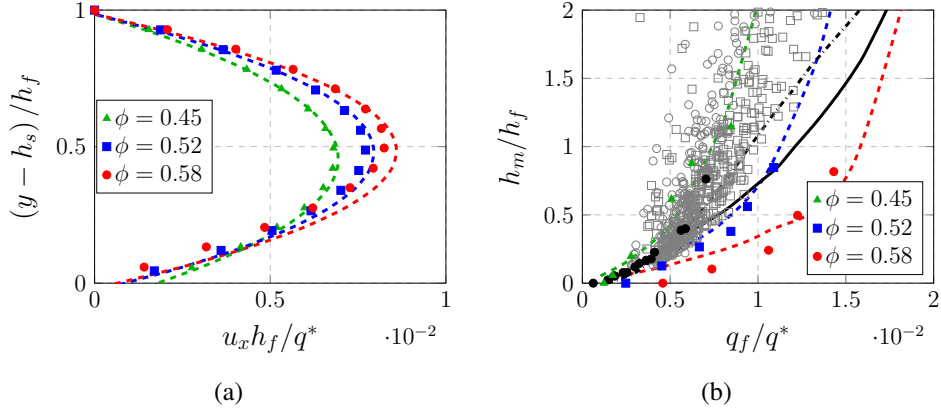}}
\label{fig:Sediment-hm-q-ana}
}
\caption{(a) Comparison of the simulation (markers) and theoretical (equation (\ref{eq-uf}), the corresponding coloured dashed lines) streamwise velocity profile in the pure fluid region when $Re = 2.16$. (b) Same as Fig. \ref{fig:Sediment-hm-q} but with the analytical solution using the parameters from the shear cell test (the corresponding coloured dashed lines). The two-phase modelling results reported by Aussillous et al. \cite{Aussillous2013} using (i) the granular rheology with $\eta_{e} / \eta_{f} = 6.6$ and $\mu (\dot{\gamma})$ (black dash-dotted line) and (ii) the suspension rheology of Boyer et al. \cite{Boyer2011} (black solid line) are included.}
\label{fig:Sediment-ana}
\end{figure}

In the final step, we extracted the fluid-sediment interface velocity under different fluid flow rates and obtained the corresponding $\eta_{e}$ and $\mu$ from the shear cell test. The two parameters were then substituted into an analytical solution derived by Aussillous et al. \cite{Aussillous2013}, which is written as,

\begin{equation}
u_{f} = \frac{\partial p / \partial x}{2 \eta_{f}} \left( y - h_{s} \right ) \left( y - L_{y} \right ) - \frac{u_{in}}{h_{f}} \left( y - L_{y} \right ),
\label{eq-uf}
\end{equation}

\noindent where $u_{in} = \left( \partial p / \partial x + \mu \phi \Delta \rho g \right) h_{m}^{2} / \left( 2 \eta_{e} \right)$ is the fluid-sediment interface velocity and $\eta_{f}$ is the dynamic viscosity of the fluid. The resultant velocity profile is shown in Fig. \ref{fig:purefluid} where good agreement between the simulation and equation (\ref{eq-uf}) is achieved.

The normalised bedload thickness is computed as,

\begin{equation}
\frac{h_{m}}{h_{f}} = \frac{\eta_{e}}{\eta_{f}} \left[ \sqrt{1 - \frac{\eta_{f}}{\eta_{e}} \frac{\partial p / \partial x}{\partial p / \partial x + \mu \phi \Delta \rho g}} -1 \right],
\label{eq-hmhf}
\end{equation}

\noindent and the normalised fluid flux is calculated as,

\begin{equation}
\frac{q_{f}}{q^{*}} = -\frac{1}{12} \frac{\partial p / \partial x}{\Delta \rho g} + \frac{\eta_{f}}{\eta_{e}} \left( \frac{\partial p / \partial x}{\Delta \rho g} + \mu \phi \right)\left[ \frac{1}{4} \left( \frac{h_{m}}{h_{f}} \right)^{2} + \frac{1 - \phi}{6} \left( \frac{h_{m}}{h_{f}} \right)^{3} \right].
\label{eq-qf-hm}
\end{equation}

Substituting (\ref{eq-hmhf}) to (\ref{eq-qf-hm}) gives the variation of $h_{m} / h_{f}$ with $q_{f} / q^{*}$ and the results are presented in Fig. \ref{fig:Sediment-hm-q-ana}. The friction coefficient for $\phi = 0.45$ suffered severe scattering due to low particle-plate collision frequency. Hence $\mu = 0.28$ was used to fit the data. By taking the non-Newtonian rheology into consideration, quadratic curves are obtained by the theoretical expression, which achieves good agreement with our simulation results. The two-phase modelling results using the granular rheology \cite{Cassar2005}, black dash-dotted line) and the suspension rheology \cite{Boyer2011}, black solid line), which were reported by Aussillous et al. \cite{Aussillous2013}, are also included in the figure. Note that these results also exhibits a nonlinear tendency. However, $\phi$ and $\eta_{e}$ were both kept constant in these models, which might be the reason for the reported underestimation of the mobile layer thickness by the two-phase modelling at large fluid flow rates.

\section{Conclusions}
\label{sec:conclusion}

We have conducted a numerical investigation on bedload thickness of the subaqueous sediment in laminar shearing flow, via the MPSM that allows a direct, accurate access to the solid volume fraction inside the sediment. Sediments with three solid volume fractions were simulated. Results have shown that for a densely-packed sediment, nonlinear relationship is obtained between the normalised thickness of the mobile layer and the normalised fluid flow rate. In an attempt to explain such quadratic relationship, a shear cell test was conducted to measure the shear-rate-dependent effective viscosity and friction coefficient of the sediment. Substitution of these two parameters into a theoretical expression successfully replicates and validates the simulation results. 

In summary, slight shear-thickening behaviour is first exhibited by the sediment at low Reynolds number. With the increase of the fluid flux and thus the interface velocity, the sediment enters the shear-thinning regime. By taking the non-Newtonian rheology into consideration, nonlinear relationship between the thickness of the mobile layer and the fluid flow rate is exhibited by the subaqueous sediment.

Despite the promising results that have been presented in this paper, improvements are still desired to accomplish a broader range of applications for the LBM-DEM-MPSM approach. For instance, the incorporation of a turbulence model might sufficiently extend the applicability of the model and better mimic the real-case sediment behaviour in turbulence. This is a potential objective for future work, along with an extension to simulate non-isothermal scenarios.

\bibliographystyle{elsarticle-num}
\biboptions{numbers,sort&compress}
\bibliography{granular_rheology_paper}

\end{document}